\documentclass[amsmath,amssymb,a4paper,twocolumn,unpublished]{quantumarticle}
\pdfoutput=1
\usepackage[colorlinks=true, linkcolor=red, citecolor=blue, urlcolor=blue]{hyperref}
\usepackage[utf8]{inputenc}
\usepackage{graphicx}
\usepackage{algorithm}
\usepackage{algpseudocode}
\usepackage{amssymb}
\usepackage{amsmath}
\usepackage{dsfont}
\usepackage{bm}
\usepackage{amsthm}
\usepackage{dsfont}
\usepackage{enumerate}
\usepackage{stmaryrd}
\usepackage[percent]{overpic}
\usepackage{mathtools}
\usepackage{upgreek}

\newcommand{\ket}[1]{\left|{#1}\right\rangle}
\newcommand{\bra}[1]{\left\langle{#1}\right|}
\newcommand{\kb}[1]{\left|#1\right\rangle\!\!\left\langle #1\right|}

\newcommand{\RN}{\mathbb{R}}
\newcommand{\CN}{\mathbb{C}}
\newcommand{\QN}{\mathbb{Q}}

\newcommand{\interp}[1]{\llbracket #1 \rrbracket}
\newcommand{\interpM}[1]{\left\llbracket #1 \right\rrbracket_{\mathcal{M}^{(B)}}}

\newtheorem{theorem}{Theorem}
\newtheorem{proposition}[theorem]{Proposition}

\newtheorem{lemma}[theorem]{Lemma}

\newcounter{continue}

\begin{document}
\title{Quantum Advantage needs Multiple Perspectives}

\author{Jochen Szangolies}
\email{Jochen.Szangolies@dlr.de}
\affiliation{Institute for Software Technology, German Aerospace Center (DLR), Cologne, Germany}
\orcid{0000-0003-1141-3225}

\begin{abstract}

The origin of the quantum speedup is a topic of ongoing research. We propose a logical criterion for the possibility of a speedup within the continuous-variable and odd-prime qudit settings. There, Wigner function nonnegativity signals the possibility of efficient classical simulation in the state-injection scenario. We consider representations of quadrature/Weyl measurements in a single von Neumann algebra that reproduce the Born statistics in a particular state. If in that state, the set-theoretic universe of the algebra is ZFC-satisfiable, i.e. every theorem of ZFC set theory holds with unit probability, all measurement contexts can be represented in a single commutative algebra of `beables'. Furthermore, the statistics admit a noncontextual hidden variable model, ensuring Wigner nonnegativity for two or more modes or qudits. For a single mode, a stronger notion of linear representation ensures commutativity, and with it, ZFC-satisfiability independently of the state, as well as Wigner nonnegativity. In consequence, in any setting that cannot be efficiently simulated, it must be impossible to merge the models associated with every context into a single ZFC-satisfiable one: quantum advantage needs multiple perspectives.

\end{abstract}

\maketitle

\section{Introduction}

The application of mathematical logic, and specifically model theory, to quantum mechanics started in earnest in the late 1970s with the seminal contributions of Paul Benioff~\cite{benioff1976amodels, benioff1976bmodels} and Gaisi Takeuti~\cite{takeuti1977boolean, takeuti1978two, takeuti1981quantum}. A key aspect of these works is the use of Boolean-valued models of set theory (Zermelo-Fraenkel set theory with the axiom of choice, ZFC for short). Roughly speaking (see below for the details), a Boolean-valued model of ZFC is a mathematical structure in which all theorems of ZFC hold, while at least some statements undecidable for ZFC are assigned an `intermediate' truth value $b$, with $0<b<1$ and $b$ being an element of a complete Boolean algebra $B$. 

To build some intuition, one can motivate the relevance of such objects to quantum mechanics via the notion that at least some measurement outcomes are irreducibly random. Indeed, under certain conditions, quantum mechanical measurement outcomes are $1$-random~\cite{landsman2021indeterminism}, meaning that no initial segment of an infinite outcome sequence $\sigma$ can be compressed by more than a constant factor by computational means~\cite{calude2002information}. Such a $\sigma$ cannot be computable (recall von Neumann's warning that `any one [\emph{sic}] who considers arithmetical methods of producing random digits is, of course, in a state of sin' \cite{von1963various}), for if some program $p$ were to output $\sigma$ if provided to a universal, deterministic Turing machine $U$, this would constitute compression of $\sigma$ to the length of $p$. Consequently, no formal system $T$ capable of expressing the necessary arithmetic to define $\sigma$ can at the same time decide the value of all of its bits: otherwise, there would be a program successively enumerating proofs of $T$ until, for any $n$, it found a proof that the $n$-th digit of $\sigma$ is either $1$ or $0$ (in a binary representation). But this would render $\sigma$ computable, as the resulting program could produce its digits in sequence. Therefore, if quantum measurement is genuinely random, not all measurement results can be decidable for some appropriately strong formal theory $T$.

This yields the motivation to study Boolean-valued models: whenever there is an undecidable proposition $\phi$ within a (first-order) theory $T$, it follows from the completeness theorem that there are models $\mathcal{M}_1$, $\mathcal{M}_2$ of $T$ such that $\mathcal{M}_1\models\phi$ and $\mathcal{M}_2\models\neg\phi$, i.e. $\mathcal{M}_1$ and $\mathcal{M}_2$ are structures such that every theorem of $T$ holds in them, and in $\mathcal{M}_1$, $\phi$ holds, while in $\mathcal{M}_2$, so does its negation. Furthermore, in some Boolean-valued model, $\phi$ may be assigned an intermediate truth value in the above sense. Thus, Boolean-valued models are useful to study undecidable statements. 

To this end, they were introduced in work by Scott and Solovay (see, e.g., \cite{bell2011set}) to provide an alternative formulation of Cohen's notion of \textit{forcing}, as developed in his celebrated proof of the independence (undecidability) of the continuum hypothesis from the axioms of ZFC~\cite{cohen1963independence}. Cohen showed that there are models of ZFC in which the continuum hypothesis is true, and models in which it is false; hence its truth cannot be established from the axioms of ZFC.

The first significant result is Takeuti's proof that moving from an `ordinary' to a Boolean-valued model, the real numbers of that model are in one-to-one correspondence with the self-adjoint operators of a commutative sublattice on an infinite-dimensional, separable Hilbert space~\cite{takeuti1978two}. That is, this move is akin to a logical version of quantization: real values are replaced by Hilbert-space operators. Benioff~\cite{benioff1976bmodels} has then shown that, starting with a minimal model of ZFC, an infinite random outcome sequence forces a Cohen extension of the model. 

These results highlight the relevance of undecidability, or logical independence, for quantum mechanics. In~\cite{szangolies2018epistemic}, it was proposed that this has foundational significance: it is, in fact, due to the existence of undecidable measurement outcomes (for any given observer) that a quantum mechanical description is necessary. In part, we will elaborate on this in the following: undecidability necessitates the introduction of a family of Boolean-valued models for each measurement context, which cannot be subsumed into a single model, yielding a logical version of the Kochen-Specker theorem~\cite{kochen1968hidden}. Furthermore, the concept of a Boolean-valued model itself provides a rigorous formulation of the notion of an \textit{epistemic horizon} introduced in~\cite{szangolies2018epistemic}.

Our focus, however, lies on the relevance of these mathematical logical considerations for the characterization of the quantum computational speedup. To this end, our main result is to show that, in qudit-systems and in the setting of continuous-variable (CV) quantum computation, a necessary requirement for the presence of a quantum speedup is that the underlying set-theoretical model is \textbf{not} \textit{ZFC-satisfiable} in the sense that every theorem of ZFC holds with probability one. Thus, for any ZFC-satisfiable model, the system can be efficiently simulated on a classical computer.

The computational model used throughout is that of quantum computing by state injection (QCSI): enhancing a set of gates and measurements not itself capable of delivering an advantage over classical resources with certain special (e.g. `magic'~\cite{bravyi2005universal}) resource states that these means are unable to produce. This yields an ideal window into the origins of the quantum speedup: whatever characteristics pick out these states as useful for QCSI is at least a promising candidate for driving the power of quantum computation.

To keep the presentation as self-contained as possible, we will start with a brief introduction to Boolean-valued models in general, and their application to quantum mechanics specifically, in Sec.~\ref{sec:BVM}. The purpose of this section is mainly to foster intuition for a topic usually remote from most physicists' interests, and thus, builds its exposition alongside an explicit toy-example, while the reader is directed to the cited literature for technical details. A particular focus is on the work of Ozawa, most notably the logical characterization of beable-algebras advanced in~\cite{ozawa2023logical}, which forms the first key support column for our result.

The second column is then discussed in Sec.~\ref{sec:CVQC}. There, we recall some fundamentals of CV quantum computation, in particular with a view to introducing the result by Booth et al.~\cite{booth2022contextuality} (see also~\cite{haferkamp2021equivalence}) that contextuality and Wigner-function negativity are equivalent resources in this setting. Wigner nonnegativity is sufficient for the existence of an efficient classical simulation of the system~\cite{mari2012positive} (for operations themselves having a Wigner-nonnegative representation), thus, any setting admitting a noncontextual hidden variable model cannot provide a speedup. 

An analogous characterization holds for qudit-quantum computation with systems of odd prime dimension~\cite{delfosse2017equivalence}. Thus, in Sec.~\ref{sec:qudit}, we present the first run-through of our construction for qudit systems. While the most immediate strategy of trying to find a subalgebra of the operators on the system's Hilbert space that is `beable' in the sense of admitting simultaneous values does not work, an auxiliary phase-space construction yields a positive result, leading us to introduce the notion of a \textit{classical system} for a given setting. 

The construction is then extended to CV quantum systems in Sec.~\ref{sec:conti}. We follow essentially the same blueprint as in the preceding section, extending the concepts as necessary to apply to the infinite-dimensional setting. We demonstrate that the algebra of functions on phase space serves as a generalized `beable' algebra, whose existence is equivalent to the nonnegativity of the Wigner function. 

In Sec.~\ref{sec:logic}, we then connect these two constructions to the formalism developed by Ozawa~\cite{ozawa2023logical}. We show that if a classical system exists, we can construct a model in which every ZFC theorem holds, thus connecting the mathematical logical criterion of ZFC-satisfiability with the potential for the existence of a quantum speedup. Moreover, ZFC-satisfiability turns out to entail the noncontextuality of a setting, thus demonstrating that a quantum computational advantage can only obtain if the models associated with every context cannot be amalgamated into a single model satisfying every ZFC-theorem. 

Finally, we conclude and discuss some possible directions for future research in Sec.~\ref{sec:conc}.

\section{Boolean-Valued Models and Quantum Mechanics}\label{sec:BVM}

In this section, we first recall some basic notions from mathematical logic, notably model theory, in order to supply some of the background necessary to properly appreciate the application of Boolean-valued models to quantum theory. When we then introduce BVMs, we will do so by means of a simplified toy example, taken from~\cite{hoffmann2018grenzen}, sufficient to introduce most main concepts. Then, we will give a brief review of the application of BVMs to the formulation and interpretation of quantum mechanics. 

\subsection{Axioms and Models}

Intuitively, a \textit{formal theory} $T$ is given by a set of \textit{axioms} such that its \textit{theorems} can be derived using the repeated application of certain \textit{rules of inference}. A \textit{model} $\mathcal{M}$ of $T$ then is a mathematical structure (typically some set equipped with finitary operations and relations) that \textit{satisfies} the axioms in the sense that, suitably interpreted, they come out true. We may for example think of the theory of fields, which has among its models the rational numbers $\QN$, the real numbers $\RN$, the complex numbers $\CN$, or the two-element field $\mathbb{F}_2$, each equipped with the usual multiplication and addition.

More formally, we fix a \textit{formal language} $\mathcal{L}$, which is a set of logical and non-logical symbols equipped with a grammar that governs which strings of symbols correspond to \textit{well-formed formulas}. The non-logical symbols comprise the \textit{signature}, with symbols for constants as well as functions and relations of a certain arity. Here, an $n$-ary function given a set $S$ is a mapping $S^n \rightarrow S$, while an $n$-ary relation is a subset of $S^n$. Constants may be viewed as relations of arity zero. Together with variables, logical connectives, quantifiers, and other logical symbols, these form the well-formed formulas according to the grammar. Well-formed formulas without free variables are called \textit{sentences} of the theory.

A finitely specifiable subset of these sentences comprise the theory's axioms. This set need not itself be finite, but may contain one or more \textit{axiom schemas}, which contain placeholders for $\mathcal{L}$-expressions together with rules specifying admissible replacements. 

We call $T$ \textit{consistent} if there is no sentence $\phi$ such that both it and its negation $\neg\phi$ can be derived from the axioms by means of applying the rules of inference. We write $T\vdash\phi$ if $T$ proves $\phi$, that is, $\phi$ is derivable from the axioms using a finite sequence of applications of the rules of inference. If furthermore for every well-formed formula $\phi$, either $T\vdash\phi$ or $T\vdash\neg\phi$, we call $T$ \textit{complete}.

It is a straightforward consequence of the completeness theorem that for a consistent $T$ whose language is first order (that is, the quantifiers $\forall$ and $\exists$ range only over individuals of the domain), there is a model of $T$. Furthermore, for such a theory, every logical consequence of the axioms is derivable in finitely many steps. A model $\mathcal{M}$ of $T$ is a tuple $(U,I)$ consisting of a set $U$ (the \textit{universe} of the model) together with an \textit{interpretation} $I$ that takes any function or relation symbol to a function or relation defined over $U$. We write that $\mathcal{M}\models\phi$ (read `$\mathcal{M}$ models $\phi$') if under this interpretation, $\phi$ is true in $\mathcal{M}$.

In general, a particular $T$ need not be complete, i.e. there exist formulas $\phi$ such that neither $T\vdash\phi$ nor $T\vdash\neg\phi$. As an example, take the formula $\tau:\forall a\exists b: b\cdot b=a$ in the language of the theory of fields. It asserts the existence of a square root for every element. This formula cannot be proven from the field axioms: it is \textit{independent} of the theory. There are two possible ways to accommodate such a situation. On the one hand, one can extend the field axioms either by $\tau$ or $\neg\tau$, and obtain a consistent system in either case. By completeness, there must be models of both extensions, and indeed there are: square roots for all elements exist in $\CN$ and $\mathbb{F}_2$, but not in $\QN$ or $\RN$. 

On the other hand, one can take a particular model of $T$, and extend it by adding a new element. Thus, take the real numbers $\RN$, and add the imaginary unit $i$, and with it, square roots for all elements. The resulting structure, $\CN$, remains a model of $T$, in which however $\tau$ now holds. We can write this more succinctly using the \textit{valuation} 
\begin{align*}
    \interp{\phi}_{(U,I)}:=\left\{\begin{matrix} 1&\text{ if } (U,I)\models\phi \\0&\text{ if } (U,I)\not\models\phi \end{matrix} \right.
\end{align*}
That is, e.g. $\interp{\tau}_\RN = 0$, while $\interp{\tau}_\CN = 1$ (in a slight abuse of notation, we will sometimes refrain from explicitly noting the interpretation $I$, and take it to be understood that we replace the free variables of a formula with those elements of the model that $I$ assigns to them). In the sequel, we will be chiefly concerned with the extension of models by adding new elements.

Consequently, for a proposition independent of $T$, there will exist models of $T$ in which it holds, or fails to. Logical independence thus implies the existence of models that are not isomorphic, and indeed fail to be elementarily equivalent, just as $\CN$ contains elements that have no counterpart in $\RN$.

However, note that the converse is not true: even complete systems need not have unique models up to isomorphism, a property known as \textit{categoricity}. In fact, if $T$ has an infinite model, the L\"owenheim-Skolem theorem implies that $T$ must have models of any infinite cardinality. While this prevents a unique (up to isomorphism) model, one might hope that there are theories that have the property that for any sentence $\phi$, either it or its negation holds in every model (suitably interpreted). That this is not the case for at least a wide class of theories able to express a certain fragment of ordinary arithmetic (known as Robinson arithmetic~\cite{robinson1950essentially}) is the consequence of G\"odel's first incompleteness theorem~\cite{godel1931formal}: in the semantic formulation, for any such theory $T$, if it is consistent, there exists a sentence $\phi_G$ such that there are models $\mathcal{M}$, $\mathcal{M}^\prime$ of $T$ with $\mathcal{M}\models\phi_G$, $\mathcal{M}^\prime\models\neg\phi_G$.

This situation naturally invites the question: given a theory $T$, which $\phi$ are independent of its axioms? To address it, in particular with respect to proving the independence of the continuum hypothesis from the axioms of ZFC, Paul Cohen developed the technique of \textit{forcing} in 1963~\cite{cohen1963independence}. In brief, the aim is to show that for a theory $T$ and a sentence $\phi$, there exist models of both $\phi$ and $\neg\phi$; by the completeness theorem, this entails that $\phi$ is independent of $T$. A particularly elegant way of understanding forcing was then developed by Dana Scott and Robert Solovay in the form of Boolean-valued models (see~\cite{bell2011set}).

\subsection{Boolean-Valued Models and Forcing}\label{subsec:BVMforce}

The basic idea of a Boolean-valued model is to let the codomain of the valuation $\interp{\cdot}$ range over the elements of a general Boolean algebra (rather than over the two-element one). For the following, it will be useful to consider a toy example, largely adapted from \cite{hoffmann2018grenzen}, to build intuition. Thus, consider the three-element set $S_3=\{1,2,3\}$ and its powerset $\mathcal{P}(S_3)$. To make $\mathcal{P}(S_3)$ into a Boolean algebra $B$, we define logical operations as follows:
\begin{align*}
    a\wedge b&:= a \cap b \\
    a\vee b&:= a\cup b \\
    \bigwedge_{a\in A\subseteq B} a &:= \bigcap_{a\in A\subseteq B} a\text{ (meet)}\\
    \bigvee_{a\in A\subseteq B} a &:= \bigcup_{a\in A\subseteq B} a\text{ (join)}\\
    \neg a &:= \mathcal S_3\setminus a
\end{align*}

With this definition, $B$ is a complete Boolean algebra (the join and meet of arbitrary elements is in $B$), equipped with a partial order relation $a\le b :\Leftrightarrow a\wedge b = a$, which yields a logical implication $a\rightarrow b :\Leftrightarrow a\le b$. Under this order, the top element $\{1,2,3\}$ and the bottom element $\emptyset$ of $B$ will be referred to as $1$ and $0$, respectively.

An important notion in the following is that of a \textit{filter}. A filter $F$ on $B$ is a subset of $B$ such that
\begin{align*}
    &1\in F, 0\notin F,\\
    &a\in F,b\in B,b\ge a\Rightarrow b\in F,\\
    &a,b\in F\Rightarrow a\wedge b \in F.
\end{align*}
If furthermore $a\in F \Leftrightarrow \neg a\notin F$, then \textit{F} is called an \textit{ultrafilter}. For a subset $D$ of $B$, $D$ is called \textit{dense} if $\forall b\in B\setminus\{0\}\exists d\in D:d\le b$. Finally, $F$ is \textit{generic} if for any dense subset $D$ of $B\setminus\{0\}$, $F\cap D\ne \emptyset$. 

\begin{figure}[h] 
 \centering
 \begin{overpic}[width=0.5\textwidth]{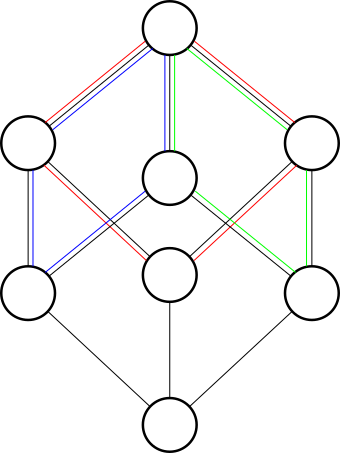}
  \put(32.8,93){$\{1,2,3\}$}
  \put(36.5,5.4){$\emptyset$}
  \put(2.8,67.5){$\{1,2\}$}
  \put(4.1,34.5){$\{1\}$}
  \put(65.5,67.5){$\{2,3\}$}
  \put(66.8,35){$\{3\}$}
  \put(34.1,60.3){$\{1,3\}$}
  \put(35.3,38.5){$\{2\}$}
  \put(55,85){\color{red} $F_2$}
  \put(1,50){\color{blue} $F_1$}
  \put(70,50){\color{green} $F_3$}
 \end{overpic}
\caption{The elements of the Boolean algebra $B=\mathcal{P}(\{1,2,3\})$. Black links between elements read in the upward direction indicate the partial ordering/implication relation, while the colors highlight the individual ultrafilters defined on $B$.}
\label{pic:bool}
\end{figure}

With this definition, there exist three distinct ultrafilters on $B$ (see also Fig.~\ref{pic:bool}):
\begin{align*}
    F_1& :=\left\{\{1,2,3\},\{1,2\},\{1,3\},\{1\}\right\}\\
    F_2&:=\left\{\{1,2,3\},\{1,2\},\{2,3\},\{2\}\right\}\\
    F_3&:=\left\{\{1,2,3\},\{1,3\},\{2,3\},\{3\}\right\}\\
\end{align*}

$\mathcal P(S_3)$, like every finite Boolean algebra, is atomic (contains minimal non-zero elements). In this case, every dense set contains the atoms and every ultrafilter is generated by a single atom (making it a \textit{principal} ultrafilter), and thus, is generic. 

We can then use $B$ to introduce the idea of \textit{Boolean sets}, where a Boolean set $S$ over $B$ is defined by assigning each member of $S$ an element of $B$ that describes its `degree of membership' in $S$. Boolean sets are defined recursively: each member is itself a Boolean set. As an example, consider a quantity that can take one of three values, which we can represent as the von Neumann ordinals $0:=\emptyset$, $1:=\{\emptyset\}$, $2:=\{\emptyset,\{\emptyset\}\}$. Since each is the set of its predecessors, the representation of such a quantity via Boolean sets needs only to record where its value exceeds $0$ or $1$ respectively:
\begin{align*}
x &:= \{(\emptyset,\{2,3\}),\ (\{(\emptyset,\{1,2,3\})\},\{3\})\},\\
y &:= \{(\emptyset,\{1,2,3\}),\ (\{(\emptyset,\{1,2,3\})\},\emptyset)\}.
\end{align*}
Each of these has two elements, which are themselves Boolean sets:
\begin{align*}
    \begin{matrix}
        x:&\emptyset \mapsto \{2,3\}, & \{(\emptyset,\{1,2,3\})\} \mapsto \{3\}\\
        y:&\emptyset \mapsto \{1,2,3\}, & \{(\emptyset,\{1,2,3\})\} \mapsto \emptyset
    \end{matrix}
\end{align*}

We can interpret this by means of a `many worlds'-kind of picture. If the elements of the base set $S_3 = \{1,2,3\}$, i.e. the atoms of the Boolean algebra, each indicate a separate `world' or possible state of affairs, then we have:

\begin{center}
    \begin{tabular}{c|c|c|c}
        & World 1 & World 2 & World 3\\ \hline
        $x$ & $\emptyset = 0$ & $\{\emptyset\}=1$ & $\{\emptyset,\{\emptyset\}\}=2$\\
        $y$ & $\{\emptyset\}=1$ & $\{\emptyset\}=1$ & $\{\emptyset\}=1$\\
        relation & $x \in y$ & $x = y$ & $y \in x$
    \end{tabular}
\end{center}

The element $(\emptyset,\{2,3\})$ of $x$ says that $0$ is an element of $x$ in worlds $2$ and $3$. The element $(\{(\emptyset,\{1,2,3\})\},\{3\})$ needs a little more unpacking: $\{(\emptyset,\{1,2,3\})\}$ contains $\emptyset$ in all worlds, so evaluates to $\{\emptyset\}$ everywhere; thus, it yields $1$ in every world. The outer weight $\{3\}$ then means that $x$ contains $1$ in world $3$. Thus, $x$ contains $0$ in worlds $2$ and $3$, and $1$ in world $3$ alone. The set $y$, on the other hand, evaluates to $1$ in every world; its elements indicate that in every world, it contains $0$, while in no world, it contains $1$.

In the remainder, we will be assuming that $\mathcal{M}$ is a model of ZFC that is:
\begin{itemize}
    \item \textit{Countable}: $|\mathcal{M}| = \aleph_0$ (which must exist due to the L\"owenheim-Skolem theorem)
    \item \textit{Transitive}: if $x\in y$ and $y\in\mathcal{M}$, then $x\in\mathcal{M}$
    \item \textit{Standard}: all elements are well-founded sets (built inductively from the empty set $\emptyset$, and the symbol $\in$ denotes the usual set membership)
\end{itemize}

We then take $\mathcal{M}^{(B)}$ to be the set of Boolean sets in $\mathcal{M}$. Each set $w$ of the base model $\mathcal{M}$ has a standard representative, or name, $\check{w}:=\{(\check{u},1)|u\in w\}$ in $\mathcal{M}^{(B)}$. As an example~\cite{hoffmann2018grenzen}, the set $w=\{\emptyset,\{\emptyset\}\}$ in $\mathcal{M}$ has the standard representative $\check{w}=\{ (\emptyset,1), (\{(\emptyset,1)\},1) \}$. In particular, $\{(\emptyset,\{1,2,3\})\}$ above is the standard representative $\check 1$ of the set $1=\{\emptyset\}$, which $x$ contains in world $3$.

We can then define Boolean `truth values' for formulas:
\begin{align*}
    \interpM{x = \check k} & =\{k+1\}\\
    \interpM{x\in y} & =\{1\}\\
    \interpM{x= y} & =\{2\}\\
    \interpM{y\in x} & =\{3\}\\
    \interpM{x\subseteq y} & =\{1,2\}\\
    \interpM{y=\check 1} & =\{1,2,3\}\equiv 1
\end{align*}
Thus, $y$ has a definite value across every world, while $x$ does not.

Note that all of the propositions being assigned an element of $\mathcal P(S_3)$ within the ultrafilter determined by a particular atom will come out true in the `world' associated with that atom. So ultrafilters, in this sense, give us the possibility of assigning consistent truth values to Boolean-valued propositions---they pick out a `world' in which they hold. This remains true once we move on to non-atomic Boolean algebras: ultrafilters allow us to pick a consistent set of propositions that hold simultaneously.

For completeness, we can then recursively define the valuation $\interpM{\cdot}$ for every ZFC-formula:

\begin{align*}
    \interpM{1}&:=1\\
    \interpM{0}&:=0\\
    \interpM{m\in n}&:=\bigvee_{v\in\mathrm{dom}(n)}\left(n(v)\wedge\interpM{m=v}\right)\\
    \interpM{m\subseteq n}&:=\bigwedge_{u\in\mathrm{dom}(m)}\left(m(u)\rightarrow \interpM{u\in n} \right)\\
    \interpM{m=n}&:=\interpM{m\subseteq n}\wedge \interpM{n\subseteq m}\\
    \interpM{\neg\phi}&:=\neg\interpM{\phi}\\
    \interpM{\phi\wedge\chi}&:=\interpM{\phi}\wedge\interpM{\chi}\\
    \interpM{\phi\vee\chi}&:=\interpM{\phi}\vee\interpM{\chi}\\
    \interpM{\forall\xi\phi(\xi)}&:=\bigwedge_{m\in\mathcal{M}^{(B)}}\interpM{\phi(m)}\\
    \interpM{\exists\xi\phi(\xi)}&:=\bigvee_{m\in\mathcal{M}^{(B)}}\interpM{\phi(m)}
\end{align*}

Here, the notation $n(v)$ can be interpreted as denoting the set of all worlds such that $v$ is in $n$, $\interpM{m=n}$ is the set of all worlds such that $m=n$, $\interpM{m\in n}$ means all worlds where $n$ contains $v$ such that $v=m$, and $\interpM{m\subseteq n}$ is true in every world where every element $u$ of $m$ (in that world) is also in $n$ (in the same world). It then follows that for every ZFC-formula $\phi$, if $\phi$ is provable from the axioms of ZFC, then $\interpM{\phi} = 1$. As an example, any two ordinals are comparable, i.e. for ordinals $s,t$, either $s<t$, $s>t$, or $s=t$. Let $\operatorname{Ord}(m)$ be the formula stating that `$m$ is an ordinal', i.e. that $m$ is transitive and linearly ordered by set membership. Then we have $\interpM{\operatorname{Ord}(x)\wedge\operatorname{Ord}(y)\mapsto (x\in y\vee x=y \vee y\in x)}=1$, with each of the worlds above realizing one out of the three possible cases.

We are now ready to introduce the notion of forcing. Essentially, forcing can be understood as generating a new model $\mathcal{M}[G]$ from a Boolean-valued model in such a way that a particular $\phi$, independent of the axioms of ZFC (and hence, of possibly `intermediate' truth value in $\mathcal{M}^{(B)}$) comes out either true or false. To do so, $G$ must be a subset of $B$ that, in some way, decides which of the elements of $B$ should receive ordinary truth values, $0$ or $1$: it must generate a Boolean algebra homomorphism from $B$ to the two-element Boolean algebra. Then, if (and only if) $\interpM{\phi}\in G$, $\mathcal{M}[G]\models\phi$.

It turns out that for this to work in a consistent manner, $G$ must be an $\mathcal{M}$-generic ultrafilter on $B$ (note that $G$ is not necessarily a part of $\mathcal{M}$, but is by construction a part of $\mathcal{M}[G]$). $G$ then determines ordinary truth values for all ZFC-independent sentences, and by choosing it appropriately, we can `force' certain sentences to come out either true or false. 

To outline the process in sequence, we start with an ordinary CTM $\mathcal{M}$ of ZFC, where $\mathcal{M}\models\phi$ for all $\phi$ such that $\mathrm{ZFC}\vdash\phi$ (likewise $\mathcal{M}\models\neg\phi$ if $\mathrm{ZFC}\vdash\neg\phi$), and for $\phi$ independent of ZFC, either $\mathcal{M}\models\phi$ or $\mathcal{M}\models\neg\phi$. From there, we transition to a Boolean-valued model $\mathcal{M}^{(B)}$ such that $\interpM{\phi}=1$ if $\mathrm{ZFC}\vdash\phi$ (or $0$ if $\mathrm{ZFC}\vdash\neg\phi$), while $\interpM{\phi}=b\in B, 0\le b \le 1$ if $\phi$ is independent of ZFC. (ZFC-independent propositions may still be assigned ordinary truth values in a Boolean-valued model, but if so, there will exist distinct models that assign different truth values.)

Then, suppose we choose $G_1$, $G_2$ such that $b\in G_1$, but $b\notin G_2$. We can then construct (ordinary) models such that $\mathcal{M}[G_1]\models\phi$, while $\mathcal{M}[G_2]\not\models\phi$. This construction works essentially by a two-stage process: we first create a quotient model $\mathcal{M}^{(B)}/ G$, where elements of $\mathcal{M}^{(B)}$ are collected into equivalence classes on the basis of the relation $x\sim_Gy:\Leftrightarrow \interpM{x=y}\in G$. This model is ordinary, meaning that it has a two-valued valuation, but not standard, because $\in$ does not have the usual interpretation of set membership. The model can be turned into a standard model by means of \textit{Mostowski collapse}: the Mostowski collapse lemma ensures that if $\in$ is `set-like' in that every element of its domain is a set, and if the model $\mathcal{M}[G]$ is well-founded, we can find a unique isomorphic CTM of ZFC. Thus, the forcing-procedure can be described by the sequence:

\begin{equation*}
    \mathcal{M} \xrightarrow[\text{choose $B$}]{} \mathcal{M}^{(B)} \xrightarrow[\text{choose $G$}]{} \mathcal{M}^{(B)}/ G \xrightarrow[\text{collapse}]{} \mathcal{M}[G]
\end{equation*}

Finally, we say that an element $p\in G$ \textit{forces} $\phi$, $p\Vdash\phi$, if $p\leq\interpM{\phi}$. This is so because for all $q\in B$, if $p\leq q$ and $p\in G$, then $q\in G$: a filter $G$ contains every element above $p$ if it contains $p$.

\subsection{Applications to Quantum Mechanics}\label{subsec:BVQM}

To make contact with empirical observation, in quantum mechanics, we need at minimum a state $\psi$ (a ray in an generally infinite-dimensional Hilbert space $\mathcal{H}$) and an observable $A$ (a self-adjoint operator acting on $\mathcal{H}$). Indeed, in the following, we will take the tuple $\left(\psi,A\right)$, a \textit{measurement context}, as a stand-in for the observer in quantum mechanics. In general, after the measurement of $A$, the system will no longer be described by $\psi$, i.e. we do not require $\psi$ to be an eigenstate of $A$. We can think of $A$ as a \textit{question} asked of the world, and of $\left(\psi,A\right)$ as encoding the necessary information to, via the Born rule, derive the best possible expectation of the answer consistent with the observer's knowledge. 

In a classical world, an analogous specification of a system's state and observable, such as a phase-space point and the system's energy, yields a definite prediction: in the state $\left(p,q \right)$, the system has a particular energy $E$. In fact, all possible questions about a system admit simultaneous, definite answers. We can thus imagine collecting all of these answers into a (in general, infinite) collection. Moreover, this collection of facts about a system is the same for all observers: in this sense, there is only a single \textit{perspective} on a classical system.

The central contention of~\cite{szangolies2018epistemic} is that this cannot be universally true: in general, there must be `questions' that can be asked of a system such that a given observer cannot answer them definitely from prior knowledge. The argument in~\cite{szangolies2018epistemic} relied on an application of Lawvere's fixed-point theorem~\cite{lawvere2006diagonal} to a logical formalization of the measurement process. However, for present purposes, it is sufficient to allow for the answers to the possible infinite collection of questions to be ($1$-)random: then, the argument from the introduction ensures that not all of these answers can be obtained by an observer, at least as long as their reasoning capabilities can be described by a finitely-specified axiomatic system, such as ZFC. In the following, we shall assume that this is indeed the case. The results obtained will then serve to put this heuristic motivation onto more stable footing.

The discussion of the preceding sections then implies that the answers to at least some questions are formally independent of the observer's knowledge, and thus, undecidable for them. They correspond to propositions $\phi$ such that neither $\phi$ nor $\neg\phi$ can be deductively established by the observer. There are models of the observer's knowledge in which $\phi$ is true, and models in which it is false. This suggests the applicability of the formalism introduced above, with at least some such propositions being assigned Boolean-valued `truth values' $b\in B$ with $0\lneq b \lneq 1$. This perspective is also discussed by Kr\'ol et al. in~\cite{krol2023random}, where it is viewed as an extension of the ideas of Paterek et al.~\cite{paterek2010logical}, who show that measurement results in quantum mechanics on finite-dimensional systems are random if they correspond to propositions undecidable from the information encoded in the state of the system.

Moreover, the fact that there are elementarily inequivalent models entails the possibility of different `perspectives' that cannot be unified into one single, overarching `view from nowhere': since in general, different Boolean algebras that overlap cannot be `glued' into a single Boolean algebra, it will not be possible to assign a single Boolean-valued model to every possible $\left(\psi,A\right)$. 

With these motivating remarks, we can now proceed to sketch the application of Boolean-valued models to quantum mechanics. We start with the observation that any self-adjoint operator $A$ has a spectral decomposition
\begin{equation}
    A=\int_{\sigma(A)} \lambda dE_\lambda,
\end{equation}
where $\sigma(A)$ denotes the spectrum of $A$ and $E_\lambda$ is a projection-valued measure. We can then choose a Boolean algebra $B$ such that $E_\lambda$ takes values in $B$ for all $\lambda$. In this case, we say that $A\in (B)$, and consider $A$ to be \textit{affiliated with} $B$. Equivalently, $A$ is affiliated with $B$ if for any unitary $U$ in the commutant of $B$, $U^* AU=A$. It can then be shown that for any family of commuting observables $\{A_\alpha\}$, there exists a Boolean algebra $B$ such that $A_\alpha\in (B)$ for every $\alpha$~\cite{takeuti1978two}. 

Moreover, for the associated model $\mathcal{M}^{(B)}$, the real numbers of the model are in one-to-one correspondence with the $A_\alpha$~\cite{takeuti1978two}. The move from an ordinary model to a Boolean-valued model whose Boolean algebra is given by the projections of a commuting set of observables is a kind of quantization in the sense that real-valued quantities are represented by self-adjoint operators, a connection emphasized by Davis~\cite{davis1977relativity}. 

In a Boolean-valued model, in general, propositions will not have a `definite' truth value in the sense that truth values may differ from $0$ or $1$. This seems closely analogous to the situation in quantum mechanics, where similarly not all possible observables can be assigned definite values in a given state.

However, notice that we have been working within a single algebra of \textit{commutative} operators $\{A_\alpha\}$. While we cannot assign simultaneous values to a \textit{non-commuting} set of observables in general, due to the Kochen-Specker theorem~\cite{kochen1968hidden}, such a valuation is possible for a single context in this sense. This is the domain of \textit{contextual} hidden variable theories or \textit{modal interpretations} of quantum mechanics~\cite{clifton1996properties}. Thus, it should be possible to assign values to any scenario $(\psi, A)$, provided we consider $A$ only in combination with (a maximal set of) commuting observables. 

The right tool for this is the mechanism of forcing, as discussed above. A generic ultrafilter in this construction can be regarded as a (contextual) hidden variable assignment~\cite{boos1996mathematical, wesep2006hidden}: for an ultrafilter $G$ any proposition such that its Boolean truth value is an element of $G$ is assigned the value $1$, while every non-element is assigned $0$. The forcing step itself corresponds to the measurement process: the final, ordinary model $M[G]$ includes definite values for all observables (within the experimental context). The generic ultrafilter $G$ corresponds to a generic random real, in the sense of an infinite binary sequence, added to the underlying model~\cite{krol2023random}.

Boos~\cite[p. 84]{boos1996mathematical} highlights these connections explicitly (notation in square brackets adapted to the one used here): 
\begin{quotation}
[...] I argue that these `hidden parameters' [$G$] also induce, via the familiar collapse of the boolean universe [$\mathcal{M}^{(B)}$] to a two-valued one [$\mathcal{M}[G]$], a metatheoretic interpretation for individual outcomes of quantum measurements, and for the notorious `collapse of the wavefunction', neither of which has a precise interpretation in von Neumann's original formalism. Such arguments derive their initial rough set-theoretic plausiblity from the simple fact that random ultrafilters [$G$] do, after all, reduce `virtual', `fuzzy', [$B$]-valued measurements in [$\mathcal{M}^{(B)}$] to `real', `sharp', two-valued ones in [$\mathcal{M}[G]=\mathcal{M}^{(B)}/ G$].
\end{quotation}

Furthermore, Boos draws a connection between the `many worlds'-interpretation of Boolean-valued models, described above, and the Everettian `worlds' of quantum mechanics~\cite[Sec. 5.2]{boos1996mathematical}.

It is instructive to consider for a moment the situation as it would have been if there were no propositions such that they cannot be assigned definite truth values. In this case, every possible model would assign the same truth value to every proposition; in this sense, there is only one single such assignment---only a single perspective. The existence of undecidability induces two levels of indefiniteness:
\begin{enumerate}
    \item The level of Boolean-valued models, where in a single such model, there are propositions that are assigned truth values different from $1$ or $0$ (`indefinite' truth values). With an ultrafilter construction and forcing, this can be taken to an ordinary model where all propositions have definite truth values. The ultrafilter used in this construction can be considered a hidden variable, while the `collapse' to an ordinary model captures the realization of definite values upon measurement. This is the level of indefiniteness of a system state $\psi$ not being in an eigenstate of a set of simultaneous measurements. It is also where Boolean-valued models provide a realization of the \textit{epistemic horizons} introduced in~\cite{szangolies2018epistemic}: for any given observer, it is exactly the set of propositions having definite truth values in the model associated with $(\psi, A)$ that circumscribe their epistemic horizon. 
    \item The level of a collection of distinct `local' Boolean-valued models for each measurement context (cf.~\cite{krol2025negation}). With the general impossibility of gluing a set of overlapping Boolean algebras into a single Boolean algebra, there is no unique (Boolean) model that could cover all possible measurement contexts. In particular, this implies the nonexistence of a Boolean-algebra homomorphism into the two-element Boolean algebra, which yields the Kochen-Specker theorem \cite{kochen1968hidden}. Physically, this corresponds to the situation of incompatible measurements performed on a quantum system.
\end{enumerate}

It is possible to `amalgamate' the Boolean-valued models of multiple incompatible contexts into a single object. This necessitates generalizing the Boolean-valued model $\mathcal{M}^{(B)}$ to a structure whose formulas take value in a more general mathematical object. This turns out to be the orthomodular lattice $\mathbb{L}(\mathcal{H})$ of (non-commuting) projections on a Hilbert space $\mathcal{H}$, the `quantum logic' of Birkhoff and von Neumann~\cite{birkhoff1936logic}. This is the approach taken in the \textit{quantum set theory} of Takeuti~\cite{takeuti1981quantum}, presently developed chiefly by Masanao Ozawa (see e.g.~\cite{ozawa2017orthomodular, ozawa2021quantum} and references therein). A remarkable result here is that the `real numbers' of an `orthomodular-valued model' $\mathcal{M}^\mathbb{L}$ still are in one-to-one correspondence with the self-adjoint Hilbert space operators. 

However, attempts to straightforwardly define such a model run into the problem that in general, it can no longer be ensured that the theorems of ZFC hold in the sense that they are assigned the top element of $\mathbb{L}$. Rather, the \textit{transfer principle} proved in~\cite{ozawa2007transfer} provides a lower bound for the truth value of ZFC-formulas. More concretely, for a ZFC-provable formula $\phi(x_1,\ldots,x_n)$ containing only bounded quantifiers (a $\Delta_0$-formula), for $u_1,\ldots,u_n\in \mathcal{M}^\mathbb{L}$, we have 
\begin{equation}\label{eq:transfer}
    \interp{\phi(u_1,\ldots,u_n)}_{\mathcal{M}^\mathbb{L}}\ge \mathrm{com}(u_1,\ldots,u_n).
\end{equation}

Here, $\mathrm{com}(u_1,\ldots,u_n)$ is the logical commutator of the `$\mathbb{L}$-valued sets' $u_1,\ldots,u_n$ such that $\mathrm{com}(u_1,\ldots,u_n) = 1$ if and only if they commute~\cite{takeuti1981quantum}. Since all elements commute if we replace $\mathbb{L}$ with a Boolean algebra $B$, this recovers the fact that all ZFC-theorems hold in a Boolean-valued model.

With the above, we can introduce the main result of~\cite{ozawa2023logical}, which will inform our analysis below. First, we need the concept of a \textit{maximal beable subalgebra} as characterized by Halvorson and Clifton~\cite{halvorson1999maximal}. The notion of a `beable' goes back to Bell~\cite{bell2004speakable}, and in a coinage paralleling `observable' is intended to convey the idea of a quantity that can be assigned a well-defined value in a given setting, such that all probabilities can be considered to stem only from ignorance about this value. The question is then, given a particular measurement context $(\psi, A)$, what is the largest collection of quantities such that they can simultaneously be assigned definite values---i.e. the maximal set of beables, in this sense.

It is a natural choice to require the set of beables to be a von Neumann algebra $\mathfrak{A}$, i.e. a subalgebra of $\mathcal L(\mathcal H)$. Furthermore, $A$ should be affiliated with $\mathfrak{A}$, $A\in(\mathfrak{A})$ in the same sense as above, namely that $\mathfrak{A}$ contains all $E_\lambda$. Then, for $f(A)$ being a Borel-function, also $f(A)\in(\mathfrak{A})$~\cite{halvorson1999maximal}. 

Now, simply choosing a maximal von Neumann algebra and its affiliated observables for the setting above would not result in a unique selection. For uniqueness, we must introduce a further (third) requirement, namely, that $\mathfrak{A}$ must be such that $U^* \mathfrak{A}U = \mathfrak{A}$ for any unitary $U$ such that $U\psi = \psi$ and $U^*AU=A$. This criterion has the effect of singling out beables in accordance with a \textit{contextualized} version of the famous `reality criterion' of Einstein, Podolsky and Rosen (EPR)~\cite{einstein1935can}. According to EPR, an observable whose value can be predicted with certainty ought to be considered an `element of reality'. This entails that correlated spin measurements on entangled Bell pairs, for instance, ought to receive simultaneous values. The prior criterion realizes this, relative to a particular measurement context $(\psi, A)$~\cite{halvorson2002reconsidering}. Thus, it can be considered an implementation of Bohr's insistence that only once the measurement context is specified, can we talk meaningfully about the properties to be considered as real (cf.~\cite{howard1979complementarity}).

The criteria discussed so far suffice to find a unique maximal beable algebra for the measurement context $(\psi, A)$~\cite{halvorson1999maximal}. It is then a result of Ozawa~\cite{ozawa2023logical} that there exists a characterization of this beable algebra by means of mathematical logic. To keep the notation simple, we will denote by $\mathcal{M}^\mathfrak{A}$ the `model' of ZFC whose propositions take truth values in the lattice of projections of $\mathfrak{A}$. If in this model, for every $\phi$ such that $\phi$ is a ZFC-theorem, $\interp{\phi}_{\mathcal{M}^\mathfrak{A}}=1$, we have a Boolean-valued model, the lattice of projections, and hence $\mathfrak{A}$, is commutative. Consequently, all of its observables can be simultaneously assigned a truth value. 

To fully characterize the algebra of beables, however, we need an additional notion: we consider a proposition $\phi(u_1,\ldots,u_n)$ \textit{satisfied in a state} $\psi$, if (omitting the parameter-dependence) $\left\langle\psi,\interp{\phi}_{\mathcal{M}^\mathfrak{A}}\psi\right\rangle=1$, that is, $\psi$ lies in the range of the projection-valued truth value of $\phi$ (and consequently, $\phi$ holds with probability 1). In this case, we write $\psi\Vdash\phi$. If for any $\Delta_0$-proposition $\phi$ such that $\mathrm{ZFC}\vdash\phi$, $\psi\Vdash\phi$, then the model $\mathcal{M}^\mathfrak{A}$ is called \textit{ZFC-satisfiable in} $\psi$ (for more details, see~\cite{ozawa2023logical}). The main result of~\cite{ozawa2023logical} is then the following theorem (\cite[Thm. 7.1]{ozawa2023logical}).

\begin{theorem}\label{thm:ZFC}
The following conditions are equivalent:
    \begin{enumerate}[\qquad (i)]
        \item $\mathcal{M}^\mathfrak{A}$ is ZFC-satisfiable in $\psi$.
        \item $\mathfrak{A}$ is a beable algebra for $\psi$.
        \item $\left[X,Y\right]\psi=0$ for any $X,Y\in\mathfrak{A}$.
        \item $\ket{\psi}\!\!\bra{\psi}\le\mathrm{com}(u_1,\ldots,u_n)$ for any $u_1,\ldots,u_n\in\mathcal{M}^\mathfrak{A}$.
    \end{enumerate}
\end{theorem}

This provides a full characterization of the maximal beable algebra in the setting $(\psi, A)$ in mathematical logical terms: $\mathfrak{A}$ is a beable algebra if and only if $\psi$ satisfies all ($\Delta_0$-)ZFC theorems in the associated model $\mathcal{M}^\mathfrak{A}$. In the following, we will investigate how to apply this characterization to the setting of odd-prime dimension and continuous-variable quantum computation. A straightforward approach turns out not to work, however, we are still able to obtain the result that efficient classical simulation is possible if the model of the measurement algebra in a given resource state $\psi$ (more generally, a mixed state $\rho$) is ZFC-satisfiable. To do so, we first recall some basics of CV quantum computation.

\section{Continuous-Variable Quantum Computation and Contextuality}\label{sec:CVQC}

This section introduces some basic concepts and notational conventions from continuous-variable quantum computing, in view towards making contact with the results of Booth, Chabaud and Emeriau~\cite{booth2022contextuality}. Therefore, our presentation is closely aligned with theirs, and we refer to the cited source for proofs and further details. 

\subsection{CV-Quantum Mechanics in Phase Space}\label{subsec:CVQM}

Let $M\in\mathbb{N}^+$ be the number of qumodes of the quantum system under discussion, with its associated Hilbert space $\mathcal{H}^M=L^2(\mathbb{R}^M)$. The position and momentum operators of the $k$-th qumode are $\hat{q}_k$ and $\hat{p}_k$, respectively. Arbitrary linear combinations of these operators are called \textit{quadratures}. 

The corresponding phase space $\mathcal{X}=\mathbb{R}^{2M}$ comes equipped with a symplectic form $\Omega(x,y)=x\cdot Jy$ for any $x, y\in \mathcal{X}$ with
\begin{equation}
    J=\begin{pmatrix} 0 & \mathds{1}_M\\
    -\mathds{1}_M & 0
    \end{pmatrix}
\end{equation}
in a particular basis $(e_k,f_k)_{k=1}^M$ of $\mathcal{X}$. $\mathcal{X}$ as $\mathbb{R}^{2M}$ comes equipped with the usual Euclidean inner product $x\cdot y$. 

Any $x = \sum_{k=1}^M q_k e_k + p_k f_k$ has an associated quadrature operator $\hat{x} =\sum_{k=1}^M q_k\hat{q}_k + p_k\hat{p}_k$. It then holds that $\left[\hat{x}, \hat{y} \right]=i\Omega(x,y)\hat 1$.

We write $\operatorname{Lag}(\mathcal{X})$ for the set of Lagrangian subspaces $L$ of $\mathcal{X}$, that is, maximal subspaces on which the symplectic form vanishes and which hence are their own symplectic dual in the sense that $L^\Omega = \left\{x\in\mathcal{X}|\Omega(x,y)=0\, \forall y\in L \right\} = L$. Since the symplectic form vanishes on $L$, all associated quadratures are simultaneously measurable, and thus, the Lagrangian subspaces form \textit{measurement contexts}. All Lagrangian subspaces are of dimension $M$.

Slightly differently to Ref.~\cite{booth2022contextuality}, we will find it more convenient to work with unitaries $e^{i\hat{x}}$ rather than displacement operators $\hat{D}(x)=e^{i\widehat{Jx}}$ throughout. These obey the Weyl relation
\begin{equation}\label{eq:Weyl}
    e^{i\hat x}e^{i\hat y}=e^{-\frac i2\Omega(x,y)}\,e^{i\widehat{x+y}}.
\end{equation}

For a state $\rho$ on $L^2(\mathbb{R}^M)$, the characteristic function $\Phi_\rho:\mathcal{X}\rightarrow\mathbb{C}$ is defined via
\begin{equation}\label{eq:FTPhi}
    \Phi_\rho(x)=\operatorname{Tr}\left(\rho\,e^{i\hat x}\right).
\end{equation}

It is continuous, since $x\mapsto e^{i\hat x}$ is strongly continuous, and satisfies $|\Phi_\rho|\leq 1$ and $\Phi_\rho(0)=1$. For a finite Borel measure $\mu$ on $\mathcal X$ we write $\mathrm{FT}(\mu):=\int_{\mathcal X} e^{ix\cdot\lambda}\,d\mu(\lambda)$, extended to tempered distributions as usual. The Wigner function is the inverse Fourier transform of $\Phi_\rho$,
\begin{equation}
    W_\rho(\lambda) := (2\pi)^{-2M}\int_{\mathcal X} e^{-ix\cdot\lambda}\,\Phi_\rho(x)\,dx,
\end{equation}

so that $\Phi_\rho=\operatorname{FT}[W_\rho d\lambda]$, yielding explicitly
\begin{equation}\label{eq:WWtrafo}
    \operatorname{Tr}\left(\rho\,e^{i\hat x}\right)=\int_{\mathcal X}e^{ix\cdot\lambda}\,W_\rho(\lambda)\,d\lambda.
\end{equation}
In particular, $\int_{\mathcal X} W_\rho\,d\lambda=\Phi_\rho(0)=1$ whenever $W_\rho\in L^1$.

The Wigner function $W_\rho$ for a quantum state $\rho$ is a \textit{pseudo}-probability distribution on $\mathcal{X}$, that is, it can in general take on negative values. It is then a remarkable result of Mari and Eisert~\cite{mari2012positive} that \textit{positivity} of the Wigner function renders classical simulation efficient, in the sense that operations with positive Wigner representation (such as Gaussian circuits and homodyne measurements) cannot produce a quantum computational speedup. More accurately, for circuits with product inputs where each factor is Wigner nonnegative, subject to positive, local gates and measurements such that one can efficiently sample phase-space points according to the associated local probability distributions, efficient classical simulation in the form of sampling from the output distribution is possible. In the following, we will always understand `efficient classical simulation' according to these criteria. In this sense, Wigner negativity is a resource for CV quantum computation enabling the possibility of a quantum advantage.

\subsection{Contextuality in the Continuous Variable-Setting}

As in Ref.~\cite{booth2022contextuality}, we follow the formalism of~\cite{barbosa2022continuous} in the setup for contextuality in continuous-variable systems. 

A \textit{(quadrature) measurement scenario} is a triple $\langle \mathcal{X}, \operatorname{Lag}(\mathcal{X}), \mathcal{O}\rangle$. Here, $\mathcal{X}$ is as before $\mathbb{R}^{2M}$, understood as a set of `measurement labels' with each $x\in\mathcal{X}$ labeling a particular quadrature measurement; $\operatorname{Lag}(\mathcal{X})$ is the set of contexts, that is, the set of all sets of compatible measurements in $\mathcal{X}$, which is here just the set of all Lagrangian subspaces of $\mathcal{X}$; $\mathcal{O}$ is the set of outcome spaces $\mathcal{O}_x$ where for each $x$, $\mathcal{O}_x:=\langle \mathbb{R},\sigma \rangle$ with $\sigma$ being the Borel $\sigma$-algebra of subsets of $\mathbb{R}$.

We call a family $e=(e_L)_{L\in \operatorname{Lag}(\mathcal{X})}$, where $e_L$ is a probability measure on the space $\mathcal{O}_L:=\mathbb{R}^L$ (equipped with the product $\sigma$-algebra) for each Lagrangian (context) $L$ an \textit{empirical model}. If there exists a probability measure $q$ on $\mathcal{O}_\mathcal{X}$ such that the marginalisation of $q$ on a context $L$, $q|_L=e_L$ for all $L$, we call $e$ \textit{noncontextual}. We will be concerned with empirical models reproducing the Born statistics for measurements on a quantum system in some state $\rho$, and thus, write $e^\rho=(e^\rho_L)_{L\in \operatorname{Lag}(\mathcal{X})}$ to make the state dependence explicit, with 
\begin{align}
\begin{split}
    e^\rho_L&\left\{s\in\mathbb R^L:s(x_k)\in E_k, k\le n\right\} \\ &= \operatorname{Tr}\left(\rho P_{\hat x_1}(E_1)\cdots P_{\hat x_n}(E_n) \right) 
\end{split}
\end{align}
for all finite families $x_1,\ldots,x_n\in L$ and Borel sets $E_k$.

The main result of~\cite{booth2022contextuality} (independently derived in~\cite{haferkamp2021equivalence}) then is to show that for any $M\ge 2$, a density operator $\rho$ on $L^2(\mathbb{R}^M)$ is noncontextual for quadrature measurements if and only if its Wigner function $W_\rho$ is nonnegative (and integrable). Thus, in the setting of two or more qumodes, contextuality and Wigner negativity are equivalent resources for CV quantum computation.

\section{Quantum Advantage in Qudit Systems}\label{sec:qudit}

As a preparation to the more involved continuous-variable setting in Sec.~\ref{sec:conti}, we first investigate the case of qudits of odd prime dimension $d$. For such systems, Veitch et al. have pointed out the importance of negative quasiprobability~\cite{veitch2012negative}, and Howard et al.~\cite{howard2014contextuality} showed that contextuality and Wigner negativity are equivalent as resources for quantum computation, in the sense that Wigner nonnegativity and noncontextuality are sufficient to ensure the existence of a classically efficient simulation in a sense analogous to that of Mari and Eisert~\cite{mari2012positive}. This result was extended to multi-qudit states with odd local dimension by Delfosse et al.~\cite{delfosse2017equivalence}, using Gross' definition of the discrete Wigner function~\cite{gross2006hudson}. We will mostly follow the definitions of the former, with notational adaptations, below.

To set the scene, we choose an odd prime $d$, write $\omega := e^{2\pi i/d}$ and $2^{-1}:=(d+1)/2\in\mathbb Z_d$. The Hilbert space of the system we consider is $\mathcal H_d:=\left(\mathbb C^d \right)^{\otimes n}$ for $n\ge 1$, and the associated (discrete) phase space is $\mathcal X_d:=\mathbb Z_d^{2n}$, whose points will be written as $a=(a_x,a_z)=((a_x)_1,\ldots,(a_x)_n,(a_z)_1,\ldots,(a_z)_n))$. $\mathcal X_d$ is equipped with the usual Euclidean inner product $a\cdot b$ and the symplectic form $\Omega_d(a,b):= a\cdot J_db = a_x\cdot b_z - a_z\cdot b_x$, in each case taken modulo $d$. 

We define the single-qudit operators $X$ and $Z$ by 
\begin{align}
\begin{split}
    X\ket{j} &= \ket{j+1},\\
    Z\ket{j} &= \omega^j\ket{j},
\end{split}
\end{align}
yielding $ZX=\omega XZ$. For a phase space point $a=(a_x, a_z)$, we define the tensor products $X^{a_x}:=X^{(a_x)_1}\otimes\ldots\otimes X^{(a_x)_n}$ and $Z^{a_z}:=Z^{(a_z)_1}\otimes\ldots\otimes Z^{(a_z)_n}$. This lets us define the discrete Weyl operator associated with each phase-space point $a$ as
\begin{equation}
    T_a := \omega^{-2^{-1}a_x\cdot a_z}Z^{a_z}X^{a_x},
\end{equation}
with the finite Weyl relation
\begin{equation}\label{eq:weyld}
    T_aT_b=\omega^{-2^{-1}\Omega_d(a,b)}T_{a+b}.
\end{equation}
This yields $T_a^*=T_{-a}$, $T_a^k=T_{ka}$ and $T_a^d=T_0=\hat 1$. The $T_a$ are orthogonal for the Hilbert–Schmidt inner product, $\operatorname{Tr}(T_a^*T_b)=d^n\delta_{ab}$, and hence linearly independent.

The eigenvalues of $T_a$ are $\omega^s$ with $s\in\mathbb Z_d$. Measuring $T_a$ corresponds to measuring the spectral projections
\begin{equation}
    \Pi_a(s):=d^{-1}\textstyle\sum_{k\in\mathbb Z_d}\omega^{-ks}T_{ka},
\end{equation}
with the outcome $s$ indicating the eigenvalue $\omega^s$. $T_a$ and $T_b$ commute on any subspace such that $\Omega_d(a,b)=0$, which yields the $n$-dimensional \textit{Lagrangian subspaces} $L_d\subseteq \mathcal X_d$. These therefore form our measurement contexts.

The discrete Wigner function for the setting is then
\begin{equation}
    W_{d,\rho}(v) := d^{-2n}\sum_{a\in\mathcal X_d}\omega^{-a\cdot v}\operatorname{Tr}(\rho\,T_a),
\end{equation}
where $v \in \mathcal X_d$. By Fourier inversion, this is equivalent to
\begin{equation}\label{eq:discW}
    \operatorname{Tr}\left(\rho\,T_a\right)=\sum_{v\in\mathcal X_d}\omega^{a\cdot v}\,W_{d,\rho}(v).
\end{equation}
This definition differs from Gross' only by a linear relabelling $v\mapsto(-v_z,v_x)$, which preserves nonnegativity.

With the results quoted above, we have a strict link between noncontextuality and classical simulability for odd-prime qudits. On the other hand, Ozawa's application of quantum set theory to the characterization of hidden variable theories in~\cite{ozawa2023logical} hints at a connection between the logical criterion of ZFC-satisfiability and the possibility of a quantum speedup. 

A first approach to bridging the two topics might be to try and construct a beable algebra for the Weyl operators, such that the existence of this beable algebra certifies the noncontextuality of the associated measurement setup, and hence, the possibility of efficient classical simulation. However, this direct route is unfortunately blocked, as the following proposition shows.

\begin{proposition}\label{prop:nogod}
Let $\mathfrak{A}\subseteq\mathcal{L}(\mathcal{H}_d)$ be a von Neumann algebra and $\psi\in\mathcal{H}_d$ a unit vector. For $T_a,T_b\in\mathfrak A$ with $\Omega_d(a,b)\ne 0$, $\mathfrak{A}$ is not beable for $\psi$. 
\end{proposition}

\begin{proof}
From Eq.~\eqref{eq:weyld}, $[T_a,T_b]=\left(\omega^{-2^{-1}\Omega_d(a,b)}-\omega^{2^{-1}\Omega_d(a,b)}\right)T_{a+b}$. The scalar is nonzero, and $T_{a+b}$ is unitary, so $[T_a,T_b]\psi$ is nonzero for every $\psi$. By Thm.~\ref{thm:ZFC}(iii), this means that $\mathfrak{A}$ is not beable for $\psi$.
\end{proof}

In particular, this entails that there is no von Neumann algebra $\mathfrak{A}\subseteq\mathcal{L}(\mathcal{H}_d)$ for the full set of Weyl operators such that the model $\mathcal{M}^\mathfrak{A}$ is ZFC-satisfiable in $\psi$ in the discrete setting: there is no `classical language' in the operator algebra of the quantum system. The reason for this is that the commutator is invertible, and hence, annihilates no state. The criterion of Thm.~\ref{thm:ZFC}(iii) can never be fulfilled.

However, it turns out that this does not completely block the program sketched above: while we cannot find the requisite ZFC-satisfiable model within $\mathcal{L}(\mathcal{H}_d)$, we \textit{can} find an appropriate structure that is not composed of operators on $\mathcal{H}_d$, but yields a commutative algebra of functions such that ZFC-satisfiability of the underlying model entails Wigner nonnegativity.

Our strategy in the following will be:
\begin{itemize}
    \item First, define \textit{context algebras} $\mathfrak{A}_L$ for each context $L$.
    \item Then, define a suitable commutative algebra $\mathfrak{C}$, such that there exist maps from all $\mathfrak{A}_L$ into $\mathfrak{C}$ that allow the representation of elements affiliated with $\mathfrak{A}_L$ within $\mathfrak{C}$.
    \item Require that $\mathfrak{C}$ is equipped with a state $p$ that reproduces the Born statistics for the given measurement setting and state.
    \item Show that such a \textit{classical system} exists if and only if $W_\rho\ge 0$.
\end{itemize}

For our context algebras, we can simply choose $\mathfrak A_L:=\operatorname{span}\left\{T_a:a\in L \right\}$. Since $T_aT_b=T_{a+b}=T_bT_a$ for $a,b\in L$, this is a commutative algebra. Joint outcomes are additive, so an assignment of values $s: L\rightarrow \mathbb Z_d$ must satisfy $s(a+b) = s(a) + s(b)$. The Born distribution is then the probability measure $m_L(s)=\operatorname{Tr}\left(\rho \Pi_L(s)\right)$, where $\Pi_L(s):=d^{-n}\sum_{a\in L}\omega^{-s(a)}T_a$ is the joint projection. 

\subsection{Classical Systems in the Discrete Setting}

Let $\rho$ be a state on $\mathcal{H}_d$. We call a \textit{classical system} a triple $\left(\mathfrak{C}, p, \pi \right)$ where
\begin{enumerate}[\qquad(C1$_d$)]
    \item\label{c1d} $\mathfrak{C}$ is a commutative von Neumann algebra and $p$ a \textit{faithful normal state} on $\mathfrak{C}$, that is, a positive, norm-one linear functional that is completely additive on orthogonal projections and vanishes on no nonzero positive element;
    \item\label{c2d} $\pi=(\pi_L)_{L\in\operatorname{Lag}(\mathcal X^d)}$ with each $\pi_L:\mathfrak{A}_L\rightarrow \mathfrak{C}$ is a \textit{unital normal $*$-homomorphism}, that is, $\pi_L$ preserves addition, multiplication, scalar multiplication, as well as the $*$-involution, maps the identity to the identity, and is continuous;
    \item\label{c3d} \textit{(coherence)} $\pi_L(T_a)=\pi_{L'}(T_a)$ whenever $a\in L\cap L'$, that is, the map $u_a:=\pi_L(T_a)$ is well defined for all $a$;
    \item\label{c4d} \textit{(Born reproduction)} $p(u_a)=\operatorname{Tr}(\rho T_a)$ for every $L$ and every $T_a\in\mathcal A_L$;
    \item\label{c5d} \textit{(linearity)} $u_a$ satisfies $u_au_b=u_{a+b}$ for all $a,b\in\mathcal X_d$.
\end{enumerate}

In the above definition, (C\ref{c5d}$_d$) is essentially the much-criticized additivity assumption of von Neumann~\cite{neumann1955mathematical}, questioned by Hermann~\cite{hermann1935naturphilosophischen} (see \cite[Chap. 15]{crull2016grete} for an English translation) and later, Bell~\cite{bell1966problem}. It holds here by construction; that it cannot be embedded in the operator algebra of the Hilbert space of the system (Prop.~\ref{prop:nogod}) may lend support to the viewpoint that von Neumann's target was not hidden variable theories as such, but rather, an assignment of values to operators within the structure of quantum mechanics (e.g.~\cite{bub2010neumann, dieks2017neumann}).

The definition is forward-looking: for a finite-dimensional $\mathfrak A_L$, normality of $\pi_L$ is automatic, but will be used in the following Sec.~\ref{sec:conti}. 

\begin{theorem}
Let $(\mathfrak C,p,\pi)$ be a classical system for $\rho$. For $v\in\mathcal X_d$, set
\begin{equation}\label{eq:proj}
    P_v:=d^{-2n}\sum_{a\in\mathcal X_d}\omega^{-a\cdot v}\,u_a\ \in\mathfrak C .
\end{equation}
Then the $P_v$ are mutually orthogonal projections with $\sum_vP_v=\hat 1$ and $u_a=\sum_v\omega^{a\cdot v}P_v$, and
\begin{equation}
    W_{d,\rho}(v)=p(P_v) \geq 0
\end{equation}
for every $v$.
\end{theorem}

\begin{proof}
$a\mapsto u_a$ is a homomorphism from $(\mathcal X_d,+)$ into the unitaries of $\mathfrak C$ yielding $u_a^* = u_{-a}$. Replacing $a$ by $a^\prime-b$ in Eq.~\eqref{eq:proj} yields

\begin{align}
\begin{split}
    u_bP_v &= d^{-2n}\sum_a\omega^{-a\cdot v}u_bu_a\\& \stackrel{(C\ref{c5d}_d)}{=}d^{-2n}\sum_a\omega^{-a\cdot v}u_{a+b}\\&=d^{-2n}\sum_{a^\prime}\omega^{-(a^\prime-b)\cdot v}u_{a^\prime}\\&=\omega^{b\cdot v}P_v.
\end{split}
\end{align}

It is here that linearity enters, as $a$ and $b$ may lie in different contexts. Thus:
\begin{align}
    P_wP_v &= d^{-2n}\sum_b\omega^{-b\cdot w}u_bP_v=d^{-2n}\sum_b\omega^{b\cdot(v-w)}P_v\nonumber\\&=\delta_{vw}P_v,\\
    \sum_vP_v &= d^{-2n}\sum_a\left(\sum_v\omega^{-a\cdot v}\right)u_a=u_0=\hat 1\\
    P_v^* &= d^{-2n}\sum_a\omega^{a\cdot v}u_a^*=d^{-2n}\sum_a\omega^{a\cdot v}u_{-a} =P_v.
\end{align}
With the above, $u_a$ can be given an explicit form by summing $u_aP_v = \omega^{a\cdot v}P_v$ over $v$, yielding $u_a = \sum_v\omega^{a\cdot v}P_v$. With (C\ref{c4d}$_d$), this yields $p(P_v) = d^{-2n}\sum_a\omega^{-a\cdot v}\operatorname{Tr}\left(\rho T_a \right) = W_{d,\rho}(v)$. Since $p$ is a state and $P_v=P_v^*P_v$, this is nonnegative.
\end{proof}

In the definition of a classical system, the algebra $\mathfrak C$ replaces the beable subalgebras of Halvorson and Clifton~\cite{halvorson1999maximal}. All probabilities ranging over the values of observables represented by the operators affiliated with such an algebra can be given an \textit{ignorance interpretation}, and thus, these are the beables of the system, in the sense that they can always be assumed to have a definite, but unknown, value. 

The above shows that if a classical system exists, then $W_{d,\rho}\ge 0$. However, to show that this is non-vacuous, it remains to exhibit a concrete example of such a classical system. We do so below, and show that it is in fact unique up to isomorphism.

Consider the set $S:=\left\{v:W_{d,\rho}(v)>0\right\}$, and let $\mathbb C^S$ be the algebra of complex functions on $S$ with pointwise operations and complex conjugation as involution, whose self-adjoint elements are the real functions and whose states are probability distributions on $S$. Then:

\begin{proposition}
Let $W_{d,\rho}\ge 0$. Then the triple $(\mathfrak C_{d,\rho}, p, \pi)$ with $\mathfrak C_{d,\rho}=\mathbb C^S$, $p(f):=\sum_{v\in S}W_{d,\rho}(v)f(v)$, and $\pi_L:\mathfrak A_L \rightarrow \mathfrak C_{d,\rho}$ with $\pi_L(T_a)=u_a$, $u_a(v) := \omega^{a\cdot v}$, is a classical system for $\rho$.
\end{proposition}

\begin{proof}
$\pi_L$ is well defined because the $T_a$ with $a\in L$ form a basis of $\mathfrak A_L$. It is multiplicative because $T_aT_b=T_{a+b}$ on $L$ and $u_au_b=u_{a+b}$; it is $*$-preserving because $T_a^*=T_{-a}$ and $\overline{u_a}=u_{-a}$; and it is unital because $T_0=\hat 1\mapsto u_0=1$. Hence, (C\ref{c2d}$_d$) holds. For (C\ref{c1d}$_d$), $\mathfrak C_{d,\rho}$ is commutative, $p$ is positive because $W_{d,\rho}>0$ on $S$ by construction, and it is faithful: if $p(f^*f)=0$, then $f=0$ on $S$, because every weight $W_{d,\rho}(v)$ there is positive. Normalization follows from $p(1) = \sum_{v\in S}W_{d,\rho}(v) = \sum_{v\in\mathcal X_d}W_{d,\rho}(v)=\operatorname{Tr}\left(\rho T_0 \right)=1$, where the second equality holds because, by the hypothesis that $W_{d,\rho}\ge 0$, $W_{d,\rho}$ vanishes on all points not in $S$, and the third is Eq.~\eqref{eq:discW} evaluated for $a=0$. (C\ref{c3d}$_d$) holds because $u_a$ does not refer to $L$. (C\ref{c4d}$_d$) is $p(u_a)=\sum_{v\in \mathcal X_d}W_{d,\rho}(v)\omega^{a\cdot v}=\operatorname{Tr}(\rho T_a)$, from Eq.~\eqref{eq:discW}, where again the sum over $S$ can be extended to the full $\mathcal X_d$ because $W_{d,\rho}\geq 0$. (C\ref{c5d}$_d$) holds because $a\mapsto \omega^{a\cdot v}$ is a character of all of $\mathcal X_d$, which remains multiplicative in the labels $a$ even for elements from different contexts.
\end{proof}

Thus, there is indeed a classical system, and since the above construction can be carried out for any $W_{d,\rho}\ge 0$, the existence of a classical system and Wigner nonnegativity are equivalent. Indeed, one can go further and show that, up to isomorphism, the above is the \textit{unique} minimal classical system for the qudit setting (`minimal' because one can always extend the system by additional elements that however add no empirically relevant content).

\begin{proposition}\label{prop:uniqd}
Suppose $(\mathfrak C^\prime, p^\prime, \pi^\prime)$ is a classical system for $\rho$ with $\mathfrak C^\prime$ generated by the $u_a^\prime$. Then $W_{d,\rho}\ge 0$, and $P^\prime_v\mapsto \delta_v$, the indicator of $v\in S$, is an isomorphism $\mathfrak C^\prime \cong \mathfrak C_{d,\rho}$ taking $u_a^\prime=\sum_v\omega^{a\cdot v}P_v$ to $u_a$ and $p^\prime$ to $p$.
\end{proposition}

\begin{proof}
The $P^\prime_v$ are orthogonal projections summing to $\hat 1$, and every $u^\prime_a$ is a combination of them, so $\mathfrak C^\prime=\operatorname{span}\{P^\prime_v\}$. Since $p^\prime$ is faithful and $p^\prime(P^\prime_v)=W_{d,\rho}(v)$, we have $P^\prime_v\neq0$ iff $v\in S$. Hence $P^\prime_v\mapsto\delta_v$ ($v\in S$) is a $*$-isomorphism onto $\mathbb C^S$. It sends $u^\prime_a=\sum_v\omega^{a\cdot v}P^\prime_v$ to $u_a$ and $p^\prime$ to $p$.
\end{proof}

Note that the construction above, with points $v$ of the discrete phase space as hidden variables, $W_{d,\rho}$ as the distribution, and outcomes given by $a\cdot v$, is essentially the model underlying the simulation algorithm of~\cite{veitch2012negative}. We cast it here in an axiomatized form to, in the following, investigate the properties responsible for the absence of a quantum advantage.

\subsection{Example: Qutrit Stabilizer}

The construction so far can be illustrated by a simple example on the basis of qutrit stabilizer states. First, we need to get a better picture of the set $S$, the set where $W_{d,\rho}$ is positive, starting with a bound on its cardinality. To this end, note that the $T_a$ are Hilbert-Schmidt orthogonal, with $\operatorname{Tr}(T_a^*T_b)=d^n\delta_{ab}$. Thus, the operators $d^{-n/2}T_a$ form an orthonormal basis. With $\rho$ expanded in this basis, we obtain:
\begin{equation}
    \operatorname{Tr}\rho^2=d^{-n}\sum_a\left|\operatorname{Tr}\left(\rho T_a\right)\right|^2.
\end{equation}

Summing $W_{d,\rho}^2$ over $v$ and using character orthogonality $\sum_v\omega^{(b-a)\cdot v}=d^{2n}\delta_{ab}$, we get:
\begin{equation}
    \sum_v W_{d,\rho}(v)^2 =d^{-2n}\sum_a \left|\operatorname{Tr}\left(\rho T_a\right)\right|^2 = d^{-n}\operatorname{Tr}\rho^2.
\end{equation}
For nonnegative $W_{d,\rho}$, it vanishes off $S$ and sums to 1. Applying the Cauchy-Schwarz inequality to the vectors $\left(\sum_{v\in S}W_{d,\rho}(v) \right)_{v\in S}$ and $(1)_{v\in S}$ yields
\begin{align}
\begin{split}
    1=\left(\sum_{v\in S}W_{d,\rho}(v)\right)^2 & \le|S|\sum_{v\in S}W_{d,\rho}(v)^2\\ &=|S|\,d^{-n}\mathrm{Tr}\rho^2,
\end{split}
\end{align}
and thus, 
\begin{equation}
    |S|\ge \frac{d^n}{\operatorname{Tr\rho^2}}\ge d^n.
\end{equation}

Consequently, no Wigner-nonnegative state can be concentrated on fewer than $d^n$ phase space points, so there are at least $d^n$ `worlds' in the sense of Sec.~\ref{sec:BVM} associated with every classical system. This provides a kind of minimal phase space volume that a classical, i.e. Wigner-nonnegative, state must occupy; in this sense, it is a discrete uncertainty principle (cf.~\cite[Thm. 2]{bu2025extremality}).  Equality is reached for a pure state, which is a stabilizer by Gross's discrete Hudson theorem~\cite{gross2006hudson}, with Wigner function $d^{-n}$ on an affine Lagrangian subspace. By the normalization of the Wigner function, the stabilizer states are exactly those for which $|S|=d^n$. 

For a single qutrit stabilizer, $|S| = d = 3$. Consider the qutrit stabilizer $\rho_1=\kb{1}$. For $a=(0,a_z)$, $T_a=Z^{a_z}$ and $\bra{1}Z^{a_z}\ket{1}=\omega^{a_z}$; for $a_x\neq 0$, $T_a\ket{1}$ is proportional to $\ket{1+a_x}$, which is orthogonal to $\ket{1}$. This yields $\operatorname{Tr}(\rho_1 T_a)=\delta_{a_x,0}\omega^{a_z}$, with which the discrete Wigner function is given by
\begin{equation}
    W_{3,\rho_1}(v) = \frac{1}{9}\sum_{a_z\in\mathbb{Z}_3}\omega^{a_z(1-v_z)}
  =\frac{1}{3}\,\delta_{v_z,1}.
\end{equation}

Thus $S=\{v_1,v_2,v_3\}$ with $v_k:=(k-1,1)$, which yields the classical system with $\mathfrak C_{3,\rho_1}=\mathbb C^S\cong\mathbb C^3$ with the uniform state $p$. Its lattice of projections is exactly the Boolean algebra $\mathcal P(\{1,2,3\})$, the example from Sec.~\ref{subsec:BVMforce}, with each atom $\{k\}$ being the indicator of $v_k$. 

The discrete geometry $\mathbb Z_3^2$ contains four lines through the origin, the two axes passing through $(0,1)$ and $(1,0)$, the line of slope $1$ given by $(1,1)$, and the line of slope 2 by $(1,2)$. These define the Lagrangian subspaces, corresponding (in the notation of Sec.~\ref{subsec:BVQM}) to the measurement contexts $(\ket{1},A)$ with $A$ referring to the spectral projections of $Z,X,XZ,XZ^2$ respectively. The Boolean truth value of the proposition `measuring $T_a$ yields $s$' is given by
\begin{equation}
    \pi_L\left(\Pi_a(s)\right)(v)=\frac{1}{3}\sum_{k\in\mathbb Z_3}\omega^{k(a\cdot v-s)}=\delta_{s,\,a\cdot v},
\end{equation}
or in the notation of Sec.~\ref{sec:BVM},
\begin{equation}
    \interp{T_a\mapsto s}=\{k:a\cdot v_k=s\}\subseteq \{1,2,3\}.
\end{equation}

For $a=(0,a_z)$, $a_z\ne 0$ we have $a\cdot v_k=a_z$ for all $k$, so the outcome $s=a_z$ has the Boolean truth value $1=\{1,2,3\}$, and the other two outcomes have truth value $0=\emptyset$. Thus, $\rho_1$ is an eigenstate of the context defined by the line through $(0,1)$, i.e. of the $Z$-operator, with $s=1$. For points $a=(1,a_z)$, $a\cdot v_k=k-1+a_z$, so the outcome $s$ has as its Boolean truth value the atom $\{k\}$ with $k\equiv s-a_z+1 \pmod 3$, with the Born probability $\frac{1}{3}$. The remaining elements of $\mathcal P(\{1,2,3\})$, e.g. $\{1,2\}$, supply the truth values of disjunctive propositions like `measuring $X$ yields $0$ or $1$'.

Each atom $k$ determines a world by means of the three ultrafilters $F_1,F_2,F_3$ of Fig.~\ref{pic:bool}. In world $k$, $Z$ has the value $1$,  $T_{(1,0)}=X$, $T_{(1,1)}=\omega^{-1}XZ$ and $T_{(1,2)}=\omega XZ^2$ have the values $k-1$, $k$ and $k+1$ (mod 3). The Born probabilities can be interpreted as ignorance about which world obtains.

With the above, we can recognize that the example of Sec.~\ref{subsec:BVMforce} was secretly about a qutrit in the state $\rho_1$ all along. To see this, define for each Weyl operator $T_a$ the name $\tau_a:=\{(\check 0,\Pi_a(1)+\Pi_a(2)),(\check 1,\Pi_a(2))\}$, recording where its value exceeds $0$ and where it exceeds $1$. The Boolean sets $x$ and $y$ are then given by $x=\pi_{L_X}(\tau{(1,0)})$ and $y=\pi_{L_Z}(\tau{(0,1)})$, where $L_X\ni(1,0)$ and $L_Z\ni(0,1)$. Then $x$ is just the value of $X$, while $y$ is the value of $Z$. $\interp{x = \check s}$ is equal to $\interp{T_{(1,0)}\mapsto s}$, and $y$ is definite because $\rho_1$ is an eigenstate of $Z$. $\interp{x\subseteq y}=\{1,2\}$ is exactly the proposition `measuring $X$ yields 0 or 1'. 

\section{Quantum Advantage in Continuous-Variable Systems}\label{sec:conti}

The previous section has established the blueprint we will now follow. Our aim is to show that the results obtained there for the discrete case carry over to the continuous variable case, with suitable adaptations. Thus, we first prove the continuous variable analogue of Prop.~\ref{prop:nogod} for the quadrature measurement setting.

\begin{proposition}\label{prop:nogo}
Let $\mathfrak{A}\subseteq\mathcal{L}(\mathcal{H})$ be a von Neumann algebra and $\psi\in\mathcal{H}$ a unit vector. For quadratures $\hat{x}$, $\hat{y}$ affiliated with $\mathfrak{A}$ and $\Omega(x,y)\ne 0$, $\mathfrak{A}$ is not beable for $\psi$. 
\end{proposition}

\begin{proof}
For $\hat{x}$ affiliated with $\mathfrak{A}$, it holds that $U\hat{x}U^*=\hat{x}$ for any unitary $U\in \mathfrak{A}^\prime$, the commutant of $\mathfrak{A}$. Then $e^{is\hat{x}}$ commutes with every unitary in $\mathfrak{A}^\prime$, hence with the whole of $\mathfrak{A}^\prime$ (since every element is a linear combination of unitaries), and thus, by the double-commutant theorem, is an element of $\mathfrak{A}$. Consider now $U_s=e^{is\hat x},\,V_t=e^{it\hat y}\in\mathfrak{A}$, which by the Weyl relation yields $U_sV_t=e^{-ist\Omega(x,y)}V_tU_s$. Consequently, 
\begin{equation}
    [U_s, V_t]\psi = \left(e^{-ist\Omega(x,y)}-1 \right)V_tU_s\psi,
\end{equation}
which is nonzero whenever $st\Omega(x,y)\notin 2\pi\mathbb{Z}$. By Thm.~\ref{thm:ZFC}(iii), this means that $\mathfrak{A}$ is not beable for $\psi$.
\end{proof}

Thus, the embedding of a beable algebra that has noncommuting quadratures as affiliated operators into the operator algebra over $\mathcal H$ is blocked in a similar way as in the discrete case (Prop.~\ref{prop:nogod}). Luckily, the same strategy introduced there can also be brought to bear on the continuous-variable setting.

For our context algebras $\mathfrak{A}_L$ we choose $\mathfrak{A}_L = W^*\left(\left\{\hat{x}:x\in L \right\}\right)$, by which we mean the smallest von Neumann algebra with which every quadrature within a given context is affiliated. Equivalently, this is the von Neumann algebra generated by the spectral projections of the quadratures. The $\mathfrak A_L$ are isomorphic to $L^\infty\left(\mathbb R^M\right)$ and maximally abelian.

\subsection{Classical Systems in the Continuous Setting}

Let $\rho$ be a state on $\mathcal{H}=L^2(\mathbb{R}^M)$. In the CV setting, we call a \textit{classical system} a triple $\left(\mathfrak{C}, p, \pi \right)$ where
\begin{enumerate}[\qquad(C1)]
    \item\label{c1} $\mathfrak{C}$ is a commutative von Neumann algebra and $p$ a \textit{faithful normal state} on $\mathfrak{C}$, that is, a positive, norm-one linear functional that is completely additive on orthogonal projections and vanishes on no nonzero positive element;
    \item\label{c2} $\pi=(\pi_L)_{L\in\operatorname{Lag}(\mathcal X)}$ with each $\pi_L:\mathfrak{A}_L\rightarrow \mathfrak{C}$ is a \textit{unital normal $*$-homomorphism}, that is, $\pi_L$ preserves addition, multiplication, scalar multiplication, as well as the $*$-involution, maps the identity to the identity, and is continuous with respect to the ultraweak topologies on $\mathfrak{A}_L$ and $\mathfrak{C}$;
    \item\label{c3} \textit{(coherence)} $\pi_L(P_{\hat x}(E))=\pi_{L'}(P_{\hat x}(E))$ whenever $x\in L\cap L'$ and $E\subseteq\mathbb R$ is Borel;
    \item\label{c4} \textit{(Born reproduction)} $p(\pi_L(A))=\operatorname{Tr}(\rho A)$ for every $L$ and every $A\in\mathcal A_L$;
    \item\label{c5} \textit{(linearity)} writing $u_x:=\pi_L(e^{i\hat x})$ for any $L\ni x$, well defined by Lemma~\ref{lem:class}(i) below, the mapping satisfies $u_xu_y=u_{x+y}$ for all $x,y\in\mathcal X$.
\end{enumerate}

The following Lemma establishes some useful properties of classical systems.

\begin{lemma}\label{lem:class}
Consider the triple $\left(\mathfrak{C}, p, \pi \right)$. Then:
\begin{enumerate}[\qquad(i)]
    \item if $\left(\mathfrak{C}, p, \pi \right)$ satisfies (C\ref{c1})-(C\ref{c3}), then for $x\in L\cap L'$, $\pi_L$ and $\pi_{L'}$ agree on all of $W^*(\hat x)$; in particular $u_x:=\pi_L(e^{i\hat x})$ does not depend on the choice of $L\ni x$, so (C\ref{c5}) is well posed;
    \item (C\ref{c4}) holds additionally if and only if $p(u_x)=\operatorname{Tr}\left(\rho\,e^{i\hat x}\right)$ for every $x\in\mathcal X$;
    \item if $\left(\mathfrak{C}, p, \pi \right)$ further satisfies (C\ref{c4}) and (C\ref{c5}), it follows that $x\mapsto u_x$ is a homomorphism from $(\mathcal X,+)$ into the unitary group of $\mathfrak C$, and $\Phi(x):=p(u_x)$ is a continuous positive-definite function on $\mathcal X$ with $\Phi(0)=1$.
\end{enumerate}
\end{lemma}

\begin{proof}
For $\left(\mathfrak{C}, p, \pi \right)$ as above, we have:
\begin{enumerate}[\qquad(i)]
    \item By (C\ref{c3}), $\pi_L$ and $\pi_{L^\prime}$ agree on the projections $P_{\hat x}(E)$, and the bounded Borel functions of $\hat x$ are uniform limits of simple functions of the projections, extending this agreement to $W^*(\hat{x})=\left\{f(\hat x): f\text{ bounded Borel} \right\}$ with $x\in L\cap L^\prime$. 
    \item The `$\Rightarrow$'-direction follows directly from setting $A=e^{i\hat x}$. For the `$\Leftarrow$'-direction, set $\mathfrak N:=\left(\operatorname{span}\{e^{i\hat x}:x\in L\}\right)$ (understood in the $\sigma$-weak topology). The span is a unital $*$-subalgebra: it contains $\hat 1=e^{i\hat 0}$, is closed under adjoints since $(e^{i\hat x})^*=e^{i\widehat{(-x)}}$, and is closed under products because the vanishing of the symplectic form on a Lagrangian $L$ implies $e^{i\hat x}e^{i\hat y}=e^{i\widehat{x+y}}$ for $x,y\in L$. Hence $\mathfrak N$ is a von Neumann algebra. Every $e^{i\hat x}$, $x\in L$, lies in $\mathfrak A_L$ by affiliation, so $\mathfrak N\subseteq\mathfrak A_L$; conversely $tx\in L$ for all $t$, so $e^{it\hat x}\in\mathfrak N$ and $\hat x$ is affiliated with $\mathfrak N$ by Stone's theorem, and consequently, $\mathfrak A_L\subseteq\mathfrak N$ by minimality. Thus $\mathfrak N=\mathfrak A_L$. Now $p\circ\pi_L$ and $\operatorname{Tr}(\rho\,\cdot\,)$ are both $\sigma$-weakly continuous and agree on a $\sigma$-weakly dense subspace, hence agree on $\mathfrak A_L$.
    \item Each $u_x$ is the image of a unitary under a unital $*$-homomorphism, hence is itself unitary, and $u_0=1$; with (C\ref{c5}) this is the homomorphism property, and $u_{-x}=u_x^*$. For positive-definiteness, let $c_1,\dots,c_n\in\mathbb C$, $x_1,\dots,x_n\in\mathcal X$, and put $A:=\sum_kc_ku_{x_k}$; then
    \begin{align}
    \begin{split}
        \sum_{j,k}\bar c_jc_k\,\Phi(x_k-x_j) &=\sum_{j,k}\bar c_jc_k\,p\left(u_{x_k}u_{x_j}^*\right)\\ &=p(AA^*)\geq0.    
    \end{split}
    \end{align}
    Continuity follows from $\Phi(x)=\operatorname{Tr}\left(\rho e^{i\hat x}\right)$, since $x\mapsto e^{i\hat x}$ is (strongly) continuous on all of $\mathcal X$, making $x\mapsto \operatorname{Tr}(\rho e^{i\hat x})$ continuous for trace-class $\rho$, and $\Phi(0)=\operatorname{Tr}\rho = 1$.
\end{enumerate}
\end{proof}

A remark on (C\ref{c5}): essentially, the requirement of linearity is what overcomes the obstruction posed by Prop.~\ref{prop:nogo}. Within each context $L$, it is automatic, since there, the symplectic form vanishes and the Weyl relation (Eq.~\eqref{eq:Weyl}) becomes $e^{i\hat x}e^{i\hat y}=e^{i\widehat{x+y}}$. Within the operator algebra of $\mathcal{H}$, the analogous relation is in general not fulfilled due to the presence of noncommuting elements $[\hat x,\hat y]=i\Omega(x,y)\hat 1$. Requiring it to hold for the whole classical system is both what will force $W_\rho\ge 0$ and why no such construction can sit in $\mathcal{L}(\mathcal{H})$.

With the above, we can now establish our main result, and show that the existence of a classical system entails positivity of the Wigner function. 

\begin{theorem}\label{thm:pos}
Let $\rho$ be a state on $L^2(\mathbb{R}^M)$ for which a classical system $(\mathfrak{C},p,\pi)$ exists. Then $W_\rho\ge 0$, $W_\rho\in L^1$, and $\mu_\rho:=W_\rho\,d\lambda$ is a Borel probability measure on $\mathcal{X}$ with $\Phi=\mathrm{FT}[\mu_\rho]$.
\end{theorem}

\begin{proof}
From Lemma~\ref{lem:class}(iii), $\Phi$ is continuous and positive-definite with $\Phi(0)=1$, so Bochner's theorem ensures the existence of a unique Borel probability measure $\mu_{\mathrm{cl}}$ on $\mathcal{X}$ with $\Phi=\operatorname{FT}[\mu_{\mathrm{cl}}]$. On the other hand, $W_\rho\in L^2(\mathcal{X})$ for every density operator (Moyal's identity), so it defines a tempered distribution, and by Lemma~\ref{lem:class}(ii) and Eq.~\eqref{eq:WWtrafo}, $\Phi=\mathrm{FT}[\mu_\rho]$ in the sense of tempered distributions. Since $\mu_{\mathrm{cl}}$ is a finite measure, it is a tempered distribution as well, and the Fourier transform is injective on $\mathcal{S}'(\mathcal{X})$. Here $\mathcal{S}(\mathcal{X})$ denotes the Schwartz space of smooth, rapidly decreasing functions on $\mathcal{X}=\mathbb{R}^{2M}$ and $\mathcal{S}'(\mathcal{X})$
its dual, the space of tempered distributions. Hence $\mu_{\mathrm{cl}}=W_\rho\,d\lambda$ as distributions, and therefore as measures. Thus $W_\rho\ge0$ almost everywhere, hence everywhere by continuity of $W_\rho$, and $\int_{\mathcal{X}}W_\rho\,d\lambda=\mu_{cl}(X)=1$.
\end{proof}


Again, the preceding shows that \textit{if} a classical system exists, the Wigner function is necessarily positive. However, nothing is said thereby regarding \textit{whether} such a classical system does, in fact, exist. As we have already seen, no such system can exist within $\mathcal{L}(\mathcal{H})$. Thus, we explicitly construct a classical system for the CV setting.


\subsection{The Phase Space System}\label{subsec:phase}

Assume in the following $\rho$ such that $W_\rho\ge 0$, and let $\mu_\rho$ be as in Thm.~\ref{thm:pos}. We assume $W_\rho\ge 0$, and set
\begin{align*}
    \mathfrak C_\rho & := L^\infty(\mathcal X, \mu_\rho), \\
    p & := \int (\cdot) d\mu_\rho, \\
    u_x &:= e^{ix\cdot \lambda}\in \mathfrak C_\rho
\end{align*}
with $\mathfrak C_\rho$ acting on $L^2(\mathcal X, \mu_\rho)$ by multiplication. The choice of $\mu_\rho$ rather than the Lebesgue measure in the definition of $\mathfrak C_\rho$ here serves to discard the regions where $W_\rho=0$, paralleling the definition of $S$ in the qudit case. We will call this the \textit{(classical) phase space system} of $\rho$.

For the above to yield the required classical system, $p$ must be a state, which is the case if $W_\rho\in L^1$, i.e. it is integrable. This is immediate for pure states, as there, $W_\rho\ge 0$ implies that the state is Gaussian~\cite{hudson1974wigner, soto1983wigner}. While it should be expected that physically sensible states are likewise integrable, the following Lemma shows that nonnegativity suffices:

\begin{lemma}
Let $W_\rho\ge 0$. Then also $W_\rho\in L^1$.
\end{lemma}

\begin{proof}
Let $\gamma_\varepsilon$ be the centred Gaussian probability density on $\mathbb R^{2M}$ with covariance $\varepsilon\hat 1$. Its characteristic function is $\int e^{ix\cdot\lambda}\gamma_\varepsilon(x)dx=e^{-\varepsilon|\lambda|^2/2}$. For $T$ a tempered distribution and $\nu$ a Schwartz function, the definition of the Fourier transform reads $\left\langle T,\operatorname{FT}(\nu) \right\rangle = \left\langle\operatorname{FT}(T),\nu \right\rangle$. Applied to $W_\rho$ and $\gamma_\varepsilon$, this yields
\begin{equation}
    \int W_\rho(\lambda)\,e^{-\varepsilon|\lambda|^2/2}\,d\lambda=\int\Phi(x)\,\gamma_\varepsilon(x)\,dx.
\end{equation}
For $\epsilon\rightarrow 0$, the right side tends to $\Phi(0)=1$, because $\Phi$ is bounded and continuous. Because $W_\rho\ge 0$, the left side increases to $\int W_\rho$ by monotone convergence. Hence, $\int W_\rho d\lambda = 1$.
\end{proof}

The construction works according to the following idea: each context algebra $\mathfrak A_L$ consists of the bounded functions of $M$ commuting quadratures. Each phase-space point $\lambda$ determines a value $x\cdot \lambda$ for the quadrature $\hat x$; thus, it effectively acts as a hidden variable. Any given context $L$ only sees the values on $L$, which we can take to a point of $\mathbb R^M$ once a suitable set of coordinates is fixed on $L$. If the model then sends any function of the quadratures in $L$ to a corresponding function of the coordinates of $\lambda$, we can represent the quantum system within $\mathfrak C_\rho$. 

Thus, fix $L\in \operatorname{Lag}(\mathcal X)$ and a basis $\left(e_1,\ldots,e_M \right)$ of $L$. Let
\begin{align*}
    r_L &:\mathcal{X}\rightarrow \mathbb R^M,\\
    r_L(\lambda) &:= \left(e_1\cdot \lambda,\ldots,e_M\cdot\lambda \right),
\end{align*}
so that for $x=\sum_i c_ie_i\in L$ we have $x\cdot\lambda = c\cdot r_L(\lambda)$. Let $P$ be the joint PVM of the (commuting) quadratures $\hat e_1,\ldots,\hat e_M$ on $\mathbb R^M$. The \textit{functional calculus} of an operator $\hat e$ is given by $f(\hat e) = \int f dP$ for bounded Borel $f:\mathbb R^M\rightarrow \mathbb C$. We will use the following properties (e.g.~\cite[Thm. 5.9]{schmudgen2012unbounded}):
\begin{enumerate}[\qquad (F1)]
    \item\label{F1} $f\mapsto f(\hat e)$ is linear, multiplicative and $*$-preserving, with $\mathbf 1(\hat e)=\hat 1$, and $f(\hat e)=0$ if and only if $f=0$ outside a $P$-null set, where $E$ is $P$-null if $P(E)=0$.
    \item\label{F2} For $x=\sum_ic_ie_i\in L$, $e^{i\hat x}=g_c(\hat e)$ with $g_c(s):=e^{ic\cdot s}$. This follows because $e^{i\hat x}e^{i\hat y}=e^{i\widehat{x+y}}$ whenever $\Omega(x,y)=0$, which gives $e^{i\hat x}=\prod_ie^{ic_i\hat e_i}$, together with multiplicativity in (F\ref{F1}).
    \item\label{F3} The image of the functional calculus is exactly $\mathfrak A_L$. This holds because by (F\ref{F2}), it contains all $e^{i\hat x}$, which by Lemma ~\ref{lem:class}(ii) span $\mathfrak A_L$. Conversely, every $f(\hat e)$ lies in $W^*(\hat e_1,\ldots, \hat e_M)\subseteq\mathfrak A_L$.
\end{enumerate}

In sum, this says that every element of $\mathfrak A_L$ is $f(\hat e)$ for some $f$, and $f$ is determined up to $P$-null sets.

Furthermore, we will make use of the following Lemma:

\begin{lemma}\label{lem:quad} Let $x = \sum_i c_i e_i \in L$, and put $h_c(s) := c \cdot s$ for $s \in \mathbb{R}^M$. Then:
\begin{enumerate}
  \item[(i)] $\hat{x} = \int_{\mathbb{R}^M} h_c \, dP$. Consequently, $P_{\hat{x}}(E) = P\left(h_c^{-1}(E)\right)=P(\{s:c\cdot s\in E\})$ for every Borel $E \subseteq \mathbb{R}$, and $f(\hat{x}) = (f \circ h_c)(\hat{e})$ for every bounded Borel $f : \mathbb{R} \to \mathbb{C}$.   
  \item[(ii)] The joint PVM $Q$ of $(\hat{x}, \hat{e}_1, \dots, \hat{e}_M)$ on $\mathbb{R}^{1+M}$ is $Q(E) = P\left(\iota_c^{-1}(E)\right)$, where $\iota_c(s) := (h_c(s), s)$. In particular, for every state $\rho$, the joint Born distribution $\mathrm{Tr}(\rho\, Q(\cdot))$ is concentrated on the hyperplane $\Gamma_c := \{(s_0, s) \in \mathbb{R}^{1+M} : s_0 = c \cdot s\}$. 
\end{enumerate}
\end{lemma}
 
\begin{proof}
(i) For $t \in \mathbb{R}$, we have $tx = \sum_i t c_i e_i \in L$, so (F\ref{F2}) gives $e^{it\hat{x}} = g_{tc}(\hat{e}) = \int e^{it h_c} \, dP$. Since $h_c$ is real-valued, $T := \int h_c \, dP$ is self-adjoint, and by the functional calculus $e^{itT} = \int e^{it h_c} \, dP$. Thus $\hat{x}$ and $T$ generate the same strongly continuous one-parameter unitary group, and $\hat{x} = T$ by the uniqueness part of Stone's theorem. By the transformation rule for spectral integrals~\cite{schmudgen2012unbounded}, the spectral measure of $T$ is $E \mapsto P(h_c^{-1}(E))$; integrating $f$ against it yields $f(\hat{x}) = \int f \circ h_c \, dP = (f \circ h_c)(\hat{e})$.
 
(ii) By (i), the spectral projections of $\hat{x}$ lie in the range of $P$, so $\hat{x}, \hat{e}_1, \dots, \hat{e}_M$ strongly commute and $Q$ is well defined. Since $\iota_c$ is continuous, the image $P \circ \iota_c^{-1}$, which assigns to each Borel set of $\mathbb{R}^{1+M}$ the projection that $P$ assigns to its preimage under $\iota_c$, is a PVM on $\mathbb{R}^{1+M}$. On rectangles, using (i) and the multiplicativity of $P$,
\begin{align}
\begin{split}
  Q(E_0 \times &E_1 \times \cdots \times E_M)
  = P_{\hat{x}}(E_0)\, P(E_1 \times \cdots \times E_M) \\
  &= P\left(h_c^{-1}(E_0) \cap (E_1 \times \cdots \times E_M)\right) \\
  &= P\left(\iota_c^{-1}(E_0 \times E_1 \times \cdots \times E_M)\right).
\end{split}
\end{align}
Two PVMs that agree on rectangles agree on all Borel sets, so $Q = P \circ \iota_c^{-1}$. Finally, $\iota_c^{-1}(\Gamma_c) = \mathbb{R}^M$, so $Q(\Gamma_c) = \hat{1}$ and $\mathrm{Tr}(\rho\, Q(\Gamma_c)) = 1$.
\end{proof}

We are now ready to show that there exists a classical system for every $W_\rho\ge 0$, the phase space system.

\begin{proposition}\label{prop:phase}
Let $W_\rho \ge0$. For each $L$, define
\begin{equation}
    \pi_L\left(f(\hat e) \right):= f\circ r_L \in\mathfrak C_\rho.
\end{equation}
Then $(\mathfrak C_\rho, p, \pi)$ is a classical system for $\rho$, and $\pi_L$ is a unital normal $*$-homomorphism $\mathfrak A_L\rightarrow \mathfrak C_\rho$ with $\pi_L(e^{i\hat x})=u_x$ for $x\in L$.
\end{proposition}

\begin{proof}
The basic idea for the proof is as follows. First, we establish that the model is capable of reproducing the Born probabilities. Then, we show that $\pi_L$ is well defined and normal, which entails (C\ref{c2}). The remaining conditions of the definition of a classical system can then be checked explicitly.

\textit{Step 1: reproduction of the Born statistics.} Let $m(E):=\operatorname{Tr}(\rho\,P(E))$ be the Born distribution of the context, a probability measure on $\mathbb R^M$. For $c\in\mathbb R^M$ and $x=\sum_ic_ie_i$,
\begin{align}
\begin{split}
    \int_{\mathbb R^M}e^{ic\cdot s}\,d\,(r_L)_*\mu_\rho(s)&=\int_{\mathcal X}e^{ix\cdot\lambda}\,d\mu_\rho(\lambda)\\ &=\operatorname{Tr}\left(\rho\,e^{i\hat x}\right)\\ &=\int_{\mathbb R^M}e^{ic\cdot s}\,dm(s).
\end{split}
\end{align}

The middle equality is just Eq.~\eqref{eq:WWtrafo}, and the final one is (F\ref{F2}). The measure $(r_L)_*\mu_\rho(E)$ is the \textit{push-forward measure} of $\mu_\rho$ through $r_L$, defined by $\mu_\rho\left(r_L^{-1}(E)\right)$. This yields an equality between the Fourier transforms of two probability measures on $\mathbb R^M$, meaning that the measures themselves must be equal: $(r_L)_*\mu_\rho = m = \operatorname{Tr}\left(\rho P(\cdot) \right)$. Thus, the construction reproduces the correct Born statistics.

\textit{Step 2: $\pi_L$ is a well-defined $*$-homomorphism.} Suppose $f(\hat e)=g(\hat e)$. By (F\ref{F1}), $f\neq g$ only on a $P$-null set $E$. Then $m(E)=0$, so  $\mu_\rho(r_L^{-1}(E))=0$, so $f\circ r_L=g\circ r_L$ in $L^\infty(\mathcal X,\mu_\rho)$. Hence $\pi_L$ is well defined on all of $\mathfrak A_L$, by the surjectivity of the functional calculus, (F\ref{F3}). It is linear, multiplicative, $*$-preserving and unital because composing with $r_L$ is, and because $f\mapsto f(\hat e)$ is, by (F\ref{F1}). By (F\ref{F2}), $\pi_L(e^{i\hat x})=g_c\circ r_L=e^{ic\cdot r_L(\lambda)}=e^{ix\cdot\lambda}=u_x$.

\textit{Step 3: $\pi_L$ is normal.} A positive map between von Neumann algebras is normal if it is completely additive on projections (this follows from \cite[Thm. 7.1.12]{kadison1986fundamentals} applied to $\phi\circ \pi_L$ for each normal state $\phi$ of $\mathfrak C_\rho$). Suppose $Q$ is a projection in $\mathfrak A_L$, then from $Q^2 = Q = Q^*$ we have $f^2 = f = \bar f$ outside of a set of $P$-measure zero. Thus, $f\in \{0,1\}$ almost everywhere, meaning $f = 1_E$ for $Q=P(E)$. Choose a countable, mutually orthogonal and nonzero family $P\left(E_n \right)$. (A countable family suffices by the separability of $L^2(\mathbb R^M)$.) Then $P(E_i\cap E_j)=P(E_i)P(E_j)=0$ for $i\ne j$, meaning the overlap of distinct sets is a $P$-null set. We can then produce a family of mutually disjoint sets by disjointification, $F_n:= E_n\setminus\bigcup_{k<n}E_k$, with each $F_n$ differing from $E_n$ only by a null set. Then $\sum_nP(F_n)=P(\bigcup_nF_n)$ by countable additivity of $P$, and correspondingly $\sum_n\mathbf 1_{r_L^{-1}F_n}=\mathbf 1_{r_L^{-1}\bigcup_nF_n}$ in $\mathfrak C_\rho$. So $\pi_L$ preserves the sum.

\textit{Step 4: checking the remaining conditions.}
\begin{enumerate}[\qquad (C1)]
    \item $\mathfrak C_\rho$ is commutative and $p=\int(\cdot)\,d\mu_\rho$ is faithful and normal. Normality follows since $p$ is the vector state of the constant $1\in L^2(\mathcal X,\mu_\rho)$, and faithfulness from the fact that $p(|f|^2)=0$ implies $f=0$ $\mu_\rho$-almost everywhere, hence $f=0$ in $L^\infty(\mathcal X,\mu_\rho)$.
    \item Established in step 2 and step 3 above.
    \item Let $x\in L\cap L'$. With $x=\sum_ic_ie_i$, by Lemma~\ref{lem:quad} the spectral projections of $\hat x$ are $P(\{s:c\cdot s\in E\})$, hence
    \begin{equation}
        \pi_L\left(P_{\hat x}\left(E\right)\right)=1_{\lambda:x\cdot\lambda\in E},
    \end{equation}
    which explicitly does not depend on $L$. More generally, $\pi_L(f(\hat x))=f(x\cdot \lambda)$ for every bounded Borel $f$ and every $L\ni x$. 
    \item With the equality of the measures established in step 1, we have:
    \begin{align}
    \begin{split}
        p\left(\pi_L(f(\hat e))\right)&=\int_{\mathcal X}f\circ r_L\,d\mu_\rho\\ &=\int_{\mathbb R^M}f\,dm\\ &=\operatorname{Tr}\left(\rho\,f(\hat e)\right).
    \end{split}
    \end{align}
    \item $u_xu_y=e^{ix\cdot\lambda}e^{iy\cdot\lambda}=u_{x+y}$ for all $x,y\in\mathcal X$; the element $u_x$ does not depend on $L$.
\end{enumerate}
\end{proof}

Consequently, for $W_\rho\ge 0$, the phase space system $(\mathfrak C_\rho, p, \pi)$ as constructed above forms a classical system. The construction is explicit for every $W_\rho\ge 0$, thus showing that the conditions of Wigner nonnegativity and existence of a classical system are equivalent. The model consists of the classical phase-space algebra of the physical system, together with the Wigner measure, such that the context algebras $\mathfrak A_L$ are realized as those phase-space functions that factor through $r_L$, that is, depend only on the $L$-directions. 

It is perhaps not so surprising that a construction like the above should exist for Wigner-nonnegative states. After all, Wigner negativity generally heralds the effect of quantum uncertainty, so for a setting in which the Wigner function never drops below zero, it should be expected to play no role. But this is exactly what is effected by the mapping to a commutative algebra, where all effects of noncommutativity are `factored out' due to the vanishing of the symplectic form. 

The Wigner distribution then essentially yields a classical probability distribution on the classical algebra of functions on phase space (modulo differences on null sets). It might be tempting to think that this is an end-run around the construction above, and simply use the nonnegative Wigner function as a hallmark of `classicality'. However, the case of multi-qubit systems frustrates this intuition: there, the Peres-Mermin argument~\cite{peres1990incompatible, mermin1990quantum} ensures contextuality independently of the system's state. But then, no classical systems can exist, even restricting to Clifford operations and Pauli measurements: there is no consistent representation into a commutative algebra across all contexts $L$ for a contextual system. 

In contrast to the results in~\cite{booth2022contextuality, haferkamp2021equivalence}, the above construction has no requirement that $M\ge 2$. In this sense, the existence of a classical model is a stronger condition than noncontextuality, as it implies Wigner nonnegativity also in the $M=1$ case. There, every Lagrangian subspace is a line through the origin; thus, distinct contexts only have the trivial quadrature $\hat 0$ in common. Hence, context distributions can always be combined into one joint distribution, thus a noncontextual model always exists. 

\begin{figure}[ht] 
 \centering
 \begin{overpic}[width=\columnwidth]{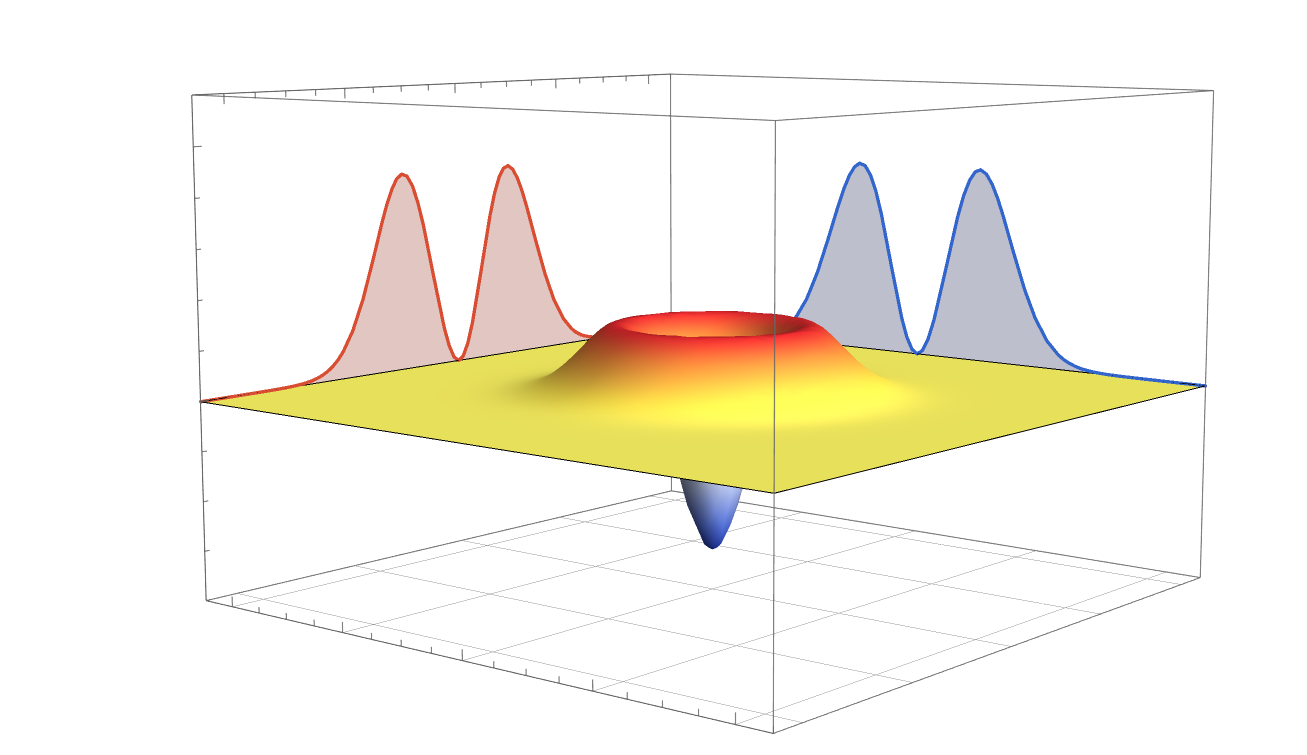}
 \put(23,10){$\mathcal X$}
 \put(33,5){$p$}
 \put(78,6){$q$}
 \put(79,41){$|\Phi(p)|^2$}
 \put(15,41){$|\Psi(q)|^2$}
 \put(6,27){$W_{\ket{1}}$}
 \end{overpic}
\caption{Graph of the Wigner function $W_{\ket{1}}$ for the Fock state $\ket{1}$, together with the associated $p$- and $q$-marginals. Since these, and those of every other context, are equal, we can assign the same distribution to each context, yielding a noncontextual hidden variable model despite the negativity of the Wigner function.}
\label{pic:Wig}
\end{figure}

In particular, for the Fock state $\ket{1}$, $W_{\ket{1}}$ is negative near the origin (see Fig.~\ref{pic:Wig}), and hence, no classical system can exist. But for quadrature measurements, a noncontextual model is possible. Parametrise the line by $\theta \in [0,\pi)$; then every quadrature $\hat x$ is of the form $t\hat q_\theta$ for real $t$. Since $\ket{1}$ is an eigenvector of the phase rotation $R(\theta)$, all $\hat q_\theta$ have the same Born distribution
\begin{equation}
    \nu (ds) = \frac{2}{\sqrt{\pi}}s^2e^{-s^2}ds.
\end{equation}

Thus for a noncontextual hidden variable model, we need simply take a real number $r$ distributed according to $\nu$, such that when measuring the quadrature $t\hat q_\theta$, the model returns $tr$, independently of $\theta$. The quantum predictions are reproduced independently of the context by a single global value assignment. However, no classical system is possible by Thm.~\ref{thm:pos}, as $W_{\ket{1}}(0)<0$. Its failure is rooted in (C\ref{c5}): in this model, quadratures $\hat q$ and $\hat p$ both receive the value $r$, but $\hat q + \hat p = \sqrt{2}\hat q_{\pi/4}$ receives $\sqrt{2}r$. This parallels Bell's criticism of von Neumann's additivity condition as applied to noncommuting operators~\cite{bell1966problem}.

The difference between the cases $M=1$ and $M\ge 2$ will be further elaborated in Sec.~\ref{sec:logic}: there, we will see that Ozawa's notion of ZFC-satisfiability in a state $\psi$ is what enables the possibility of classical simulation in the $M\ge 2$-scenario, while the linearity imposed by (C\ref{c5}) implies the stronger notion of ZFC-satisfiability in \textit{every} vector.

\subsection{Noncontextuality and \texorpdfstring{$M \ge2$}{M >= 2}}

For the $M\ge 2$ case, the above result can be achieved in an alternative way with somewhat fewer resources, by going the route via showing that a classical system entails noncontextuality, then using the results of~\cite{booth2022contextuality, haferkamp2021equivalence} to infer Wigner nonnegativity. In fact, it turns out that the classical model, in this case, need only obey (C\ref{c1})--(C\ref{c4}); the advantage of including (C\ref{c5}) in the above is then to cover also the $M = 1$ case, as well as producing a more direct route to certifying the nonnegativity of the Wigner function. 

\begin{theorem}\label{thm:noncont}
If $\rho$ admits a triple $(\mathfrak C, p, \pi)$ satisfying (C\ref{c1})--(C\ref{c4}), then there exists a probability measure $q$ on $\mathbb R^{\mathcal X}$ such that its restriction on a Lagrangian $L$, $q|_L$, equals $e^\rho_L$ for all $L$, i.e. $\rho$ is noncontextual.
\end{theorem}

\begin{proof}
For any $x\in\mathcal{X}$ and measurable Borel $E$, let $Q_x(E):=\pi_L(P_{\hat x}(E))$ for any $L\ni x$. By (C\ref{c3}), this does not depend on $L$. $Q_x$ is a projection-valued measure in $\mathfrak C$, because $\pi_L$ is a unital normal $*$-homomorphism. Since $\mathfrak C$ is commutative, all $Q_x$ commute with one another. Hence, for any finite collection $F=(x_1,\ldots,x_n)\subseteq \mathcal X$, there exists a joint projection-valued measure $Q_F$ on $\mathbb R^F$ with $Q_F(E_1\times\cdots\times E_n)=\prod_iQ_{x_i}(E_i)$~\cite{schmudgen2012unbounded}. Then set $q_F:=p\circ Q_F$, which yields a Borel probability measure on $\mathbb R^F$, which is countably additive by the normality of $p$ (see (C\ref{c1})). 

This produces a consistent family: for $F\subseteq G$, uniqueness of the joint PVM shows that the image of $Q_G$ under restriction $\mathbb R^G\rightarrow\mathbb R^F$ is $Q_F$, since both assign $\Pi_{x\in F}Q_x\left(E_x \right)$ to a rectangle, the remaining factors being $Q_x(\mathbb R) = \hat 1$. Hence the image of $q_G$ is $q_F$. 

By Kolmogorov's extension theorem, there is then a unique probability measure $q$ on $\mathbb R^{\mathcal X}$ equipped with the product $\sigma$-algebra whose image on each $\mathbb R^F$ is $q_F$. Then, for $x_1,\ldots,x_n \in L$,
\begin{align}
\begin{split}
    q\left(\{f:f(x_i)\in E_i\}\right)&=p\left(\textstyle\prod_i\pi_L\left(P_{\hat x_i}(E_i)\right)\right)\\ &=p\left(\pi_L\left(\textstyle\prod_iP_{\hat x_i}(E_i)\right)\right)\\ &=\operatorname{Tr}\left(\rho\textstyle\prod_iP_{\hat x_i}(E_i)\right),
\end{split}
\end{align}
by the multiplicativity of $\pi$ and (C\ref{c4}). Thus, under the measure $q$, each cylinder set of $\mathbb R^L$ is assigned its Born probability; but since these form a $\pi$-system generating the product $\sigma$-algebra of $R^L$, it holds in general that the image of $q$ on $\mathbb R^L$ is $e^\rho_L$ for every $L$, since two probability measures that agree on a $\pi$-system agree on the $\sigma$-algebra it generates. Thus, $\rho$ is noncontextual.
\end{proof}

(C\ref{c5}) is not used in the above proof. Consequently, for the case $M\ge 2$, the existence of a `reduced' classical system $(\mathfrak C, p, \pi)$ for $\rho$ satisfying only (C\ref{c1})--(C\ref{c4}) entails noncontextuality and hence, by the results of~\cite{booth2022contextuality, haferkamp2021equivalence},  Wigner nonnegativity. In fact, the entailment runs both directions:

\begin{proposition}\label{prop:noncont}
If $\rho$ is noncontextual for quadrature measurements, then $\rho$ admits a triple $(\mathfrak{C},p,\pi)$ satisfying (C\ref{c1})--(C\ref{c4}).
\end{proposition}

\begin{proof}
Let $q$ be a probability measure on $\mathbb{R}^{\mathcal{X}}$ (product $\sigma$-algebra) with $q|_L=e^\rho_L$ for all $L$. Set $\mathfrak{C}:=L^\infty(\mathbb{R}^{\mathcal{X}},q)$, acting by multiplication on $L^2(\mathbb{R}^{\mathcal{X}},q)$, and $p:=\int(\cdot)\,dq$; (C\ref{c1}) holds as in the proof of Prop.~\ref{prop:phase}. Write $\mathrm{ev}_x(f):=f(x)$ for the evaluation of $f$ at $x$. For each $L$, fix a basis $(e_1,\dots,e_M)$, let $r_L:=(\mathrm{ev}_{e_1},\dots,\mathrm{ev}_{e_M})$, and define $\pi_L(g(\hat e)):=g\circ r_L$. Since $q|_L=e^\rho_L$, the law of $r_L$ under $q$ is the joint Born distribution $m=\operatorname{Tr}(\rho P(\cdot))$, i.e.\ $(r_L)_*q=m$. These are all the prerequisites needed for the steps 2 and 3 of the proof of Prop.~\ref{prop:phase}, up to the explicit form of $\pi_L(e^{i\hat{x}})$, so they apply unchanged, and yield (C\ref{c2}). For (C\ref{c3}), let $x=\sum_ic_ie_i\in L$. By Lemma~\ref{lem:quad}, the joint Born distribution of $(\hat x,\hat e_1,\dots,\hat e_M)$ is concentrated on $\{s_0=\sum_ic_is_i\}$, so $\mathrm{ev}_x=\sum_ic_i\,\mathrm{ev}_{e_i}$ $q$-almost surely, and $\pi_L(P_{\hat  x}(E))=\mathbf{1}\{\mathrm{ev}_x\in E\}$, which does not depend on $L$. (C\ref{c4}) follows from $p(\pi_L(g(\hat e)))=\int g\,dm=\operatorname{Tr}(\rho\,g(\hat e))$.
\end{proof}

Note that the above has no need for $M\ge 2$, and thus shows that noncontextuality implies the existence of a reduced classical system (obeying (C\ref{c1})--(C\ref{c4})) for any number of qumodes. 

In general, depending on the distribution $q$, the Hilbert space $L^2(\mathbb{R}^{\mathcal{X}},q)$ is not necessarily separable, which is a precondition for Thm.~\ref{thm:ZFC} as stated in~\cite{ozawa2023logical}. For $M\ge 2$, there always exists a separable alternative, the $L^2(\mathcal X,\mu_\rho)$ of the phase-space system, since there, noncontextuality and Wigner nonnegativity are equivalent. The above construction for $\ket{1}$ shows a noncontextual model for the $M=1$-case on the separable $L^2(\mathbb R)$; a similar construction is possible for other states.

The above puts the present formulation in close contact with the sheaf- and topos-theoretic approaches to contextuality~\cite{abramsky2011sheaf, isham1998topos}. Recently, the relation of these frameworks to Ozawa's quantum set theory was explored in~\cite{doring2021bridge}. A detailed investigation of these connections is beyond the scope of the present work.

\subsection{Classical Surrogates and Uniqueness}

In the following section, we aim to make contact with the logical apparatus introduced in Sec.~\ref{sec:BVM}. For the discrete case, this is immediate, as the $T_a$ are directly contained in the $\mathfrak A_L$, whereas the quadratures, as unbounded operators, are only affiliated. Thus, $\mathfrak C_{d,\rho}$ already contains their images $u_a$, while the images of the quadratures within $\mathfrak C_\rho$ must be assembled from the images of their spectral projections. Thus let $(\mathfrak C, p, \pi)$ be a classical system for $\rho$. 

Since $p$ is a faithful normal state, $\mathfrak{C}$ is $*$-isomorphic, via the GNS representation of $p$, to $L^\infty(\Xi,\mu)$ acting by multiplication on $L^2(\Xi,\mu)$ for some probability space $(\Xi,\mu)$, with $p=\int(\cdot)\,d\mu$. Conditions (C1)--(C5) are preserved under $*$-isomorphism, so we can assume $\mathfrak{C}$ to be of this form without loss of generality. Self-adjoint operators affiliated with $\mathfrak{C}$ are then multiplication operators by real $\mu$-measurable functions, and we identify the two.

Then, let us define the notion of a \textit{classical surrogate} $x^\flat$ for a quadrature $\hat x$:
\begin{equation}\label{eq:surr}
    x^\flat = \int_{\mathbb{R}} tdQ_x(t).
\end{equation}

For these surrogates, we have the following:

\begin{lemma}\label{lem:affil}
$u_{tx}=e^{itx^\flat}$ for all $t\in\mathbb R$, and $x\mapsto x^\flat$ is linear.
\end{lemma}

\begin{proof}
For bounded Borel $f$, $\pi_L(f(\hat x)) = \int f dQ_x = f(x^\flat)$. For $f(s)=e^{its}$, this gives $u_{tx}=e^{itx^\flat}$. For $\mathfrak C$ acting by multiplication on an $L^2$-space, the surrogates are real measurable functions, so $e^{it x^\flat}e^{it y^\flat}=e^{it( x^\flat+ y^\flat)}$. From (C\ref{c5}), this equals $u_{t(x+y)}=e^{it(x+y)^\flat}$ for every $t$, and consequently $(x+y)^\flat = x^\flat + y^\flat$ by uniqueness from Stone's theorem. Homogeneity follows since for all $t$, $e^{it(sx)^\flat}=u_{tsx}=e^{itsx^\flat}$, with uniqueness then again yielding $(sx)^\flat = sx^\flat$.
\end{proof}

The preceding Lemma shows explicitly the connection between linearity, (C\ref{c5}), and the additivity of the $x^\flat$. With the notion of classical surrogates, we have a direct road to the continuous-variable analogue of Prop.~\ref{prop:uniqd}.

\begin{proposition}
Let $(\mathfrak C^\prime, p^\prime, \pi^\prime)$ be a classical system for $\rho$. Then there is a normal, injective, unital $*$-homomorphism $\Psi:\mathfrak C_\rho\to\mathfrak C'$ with $\Psi\circ\pi_L=\pi'_L$ for all $L$. It satisfies $p'\circ\Psi=p$, and $\mathfrak C'$ at most extends $\Psi(\mathfrak C_\rho)$ by elements in the range of no $\pi'_L$, which therefore do not affect any Born statistic.
\end{proposition}

\begin{proof}
Let $(e_k,f_k)_{k=1}^M$ be the basis of $\mathcal X$ from Sec.~\ref{subsec:CVQM}, and consider the random vector $\Lambda:=(e^\flat_k,f^\flat_k)_{k=1}^M$. Then, $x^\flat = x\cdot \Lambda$ for every $x$. From (C\ref{c4}), $p^\prime(e^{ix\cdot \Lambda})=p^\prime(u_x^\prime) = \Phi(x)$, so $\Lambda$ is distributed according to $\mu_\rho$, i.e. $p'(g(\Lambda)) = \int g \, d\mu_\rho$ for bounded Borel $g$. $\Psi$ must respect the identification in $L^\infty(\mathcal X, \mu_\rho)$, where $g$ and $g'$ are the same element if $h:= g-g' = 0$ $\mu_\rho$-almost everywhere. By linearity, it suffices to show $\Psi(h) = g(\Lambda) - g^\prime(\Lambda) = 0 \Longleftrightarrow h = 0$ $\mu_\rho$-almost everywhere. This follows by the faithfulness of $p^\prime$:
\begin{align} 
\begin{split}
  \Psi(h)=0&\Longleftrightarrow p'\left(|\Psi(h)|^2\right) = \int_{\mathcal X} |h|^2 d\mu_\rho =0 \\ &\Longleftrightarrow h = 0\quad\mu_\rho\mathrm{-a.e.}.
\end{split}
\end{align}

Moreover, $p'\circ\Psi=p$. If $h_\alpha\nearrow h$ and $k:=\sup\Psi(h_\alpha)$, then $k\le\Psi(h)$ and $p'(k)=\sup p(h_\alpha)=p(h)=p'(\Psi(h))$, so $k=\Psi(h)$ by faithfulness. Thus $\Psi$ is normal.

Since all $g(\Lambda)$ are functions of the same $\Lambda$, the map $\Psi: g \mapsto g(\Lambda)$ preserves sums, products and adjoints, and it sends $e^{ix\cdot\lambda}$ to $e^{ix^\flat} = u^\prime_x$:
\begin{equation}
\Psi\left(\pi_L(e^{i\hat x})\right)=\Psi\left(e^{ix\cdot\lambda}\right)=e^{ix\cdot\Lambda}=e^{ix^\flat}=\pi'_L(e^{i\hat x}).
\end{equation}
Thus $\Psi\circ\pi_L=\pi_L^\prime$ for every $L$, since the normal maps $\Psi\circ \pi_L$ and $\pi_L^\prime$ agree on the $\sigma$-weakly dense span of $e^{i\hat x}$, and hence on $\mathfrak A_L$.

Furthermore, $\pi'_L(\mathfrak A_L)=\Psi(\pi_L(\mathfrak A_L))\subseteq\Psi(\mathfrak C_\rho)$, so elements outside the image lie in the range of no $\pi'_L$, and $p'\circ\pi'_L=p\circ\pi_L$. Thus $\mathfrak C^\prime$ can at most extend $L^\infty(\mathcal X,\mu_\rho)$ by elements that have no empirically relevant consequences.
\end{proof}

In this sense, the phase-space system is the unique classical system for the CV case, up to isomorphism and trivial extension.

\section{Logical Characterization of Classical Simulability}\label{sec:logic}

The preceding has established the existence of classical systems as equivalent to the nonnegativity of the Wigner function, in the qudit and continuous variable-settings. Thus, by the result of Mari and Eisert~\cite{mari2012positive}, no setting admitting a classical system can provide a quantum advantage for operations with Wigner-positive representation. 

The equivalent logical characterization of classical simulability is then obtained by connecting to Ozawa's result in~\cite{ozawa2023logical}. What enables the use of Ozawa's characterization is the GNS-construction that allows the representation of a $C^*$-algebra as an algebra of bounded linear operators on an appropriate Hilbert space. Since in each case, the algebra is commutative, so is the Hilbert space algebra, and the condition of Thm.~\ref{thm:ZFC}(iii) is trivially fulfilled. Thus, the algebra $\mathfrak C$ is beable (for every vector), and the associated model $\mathcal M^{\mathfrak C}$ is ZFC-satisfiable. We call this the \textit{beable model}, to highlight the fact that the operators affiliated with the projection lattice providing its truth values are beables for the system, that is, can always be considered to have definite values, with probabilities just entering from our ignorance of these values. The dependence on the state $\rho$ enters via the existence of $\mathfrak C$ as a beable algebra for that state.

Below, we only state the results for the continuous variable-case; the discrete setting follows analogously.

\begin{proposition}\label{prop:satis}
Choose $\mathfrak C_\rho = L^\infty\left(\mathcal X,\mu_\rho \right)$ as above. Let $\mathcal H_p$ be its GNS space, such that $\mathfrak C_\rho$ is isomorphic to a subalgebra of $\mathcal L(\mathcal H_p)$. The choice of $\mathfrak C_\rho$ ensures that $\mathcal H_p$ is separable, since it acts by multiplication on the separable $L^2(\mathcal X,\mu_\rho)$~\cite[Prop. 3.4.5]{cohn2013measure}. Then we have:
\begin{enumerate}[\qquad (i)]
    \item $\mathfrak C_\rho$ is beable for any vector of $\mathcal H_p$, so by Thm.~\ref{thm:ZFC}, $\mathcal M^{\mathfrak C_\rho}$ is absolutely ZFC-satisfiable, that is, independently of the state. (The dependence on the input state is already included in the classical system.)
    \item By the Takeuti correspondence (\cite[Thm. 6.1]{ozawa2023logical}), the internal reals of $\mathcal M^{\mathfrak C_\rho}$ are exactly the self-adjoint operators affiliated with $\mathfrak C_\rho$, and thus, so are the $x^\flat$.
\end{enumerate}
\end{proposition}

The proof is immediate, with (i) following from commutativity of $\mathfrak C_\rho$, and (ii) from the definition of surrogates as affiliated operators in Eq.~\eqref{eq:surr}. Strictly speaking, for the existence of a ZFC-satisfiable model given that of a classical system, reference to Ozawa's characterization is superfluous, here: since the projection lattice of $\mathfrak C_\rho$ is commutative, the model $\mathcal M^{\mathfrak C_\rho}$ is Boolean, and thus, satisfies every ZFC theorem via the transfer principle (see Eq.~\eqref{eq:transfer}). ZFC-satisfiability, in this setting, simply reexpresses the existence of a classical system, yielding an equivalent characterization of classical simulability: every system such that a ZFC-satisfiable beable model in the above sense exists, admits of efficient classical simulation. Somewhat more stringently, we have established the following:

\begin{proposition}\label{prop:simul}
Let $(\rho_n)_n$ be a uniform family of states on $L^2(\mathbb R^{M_n})$ and $(\mathcal C_n)_n$ a uniform family of continuous-variable computations fulfilling the criteria of Mari and Eisert~\cite{mari2012positive}, i.e. Wigner-nonnegative product-state inputs subject to positive operations and measurements with efficient sampling of phase-space points from the local distributions defined by these. If every $\rho_n$ admits a classical system in its factors, or equivalently, there exists a ZFC-satisfiable beable model carrying coherent linear classical surrogates for all quadratures, then $(\mathcal C_n)_n$ admits an efficient classical simulation in the sense of efficient sampling from its output distribution.
\end{proposition}

Equivalently, a quantum advantage requires that some $\rho_n$ admits no such classical system: in other words, that there is no ZFC-satisfiable beable model for all quadratures. 

\subsection{Satisfiable Representations and the Role of Linearity}

We can use the machinery of Ozawa's characterization~\cite{ozawa2023logical} to obtain a more fine-grained understanding of the role of ZFC-satisfiability and how it applies to the question of the existence of a quantum speedup. Rather than the notion of `absolute' ZFC-satisfiability introduced above, where satisfiability holds for every $\psi$ because of the commutativity of $\mathfrak C$, Ozawa's notion of ZFC-satisfiability in $\psi$ is weaker, and does not require commutativity, but rather, only that the commutator of elements within the algebra annihilates the state (see Thm.~\ref{thm:ZFC}). In this subsection, we study what this notion entails for the possibility of classical simulation. The answer depends on whether linearity, (C\ref{c5}), is imposed: without it, ZFC-satisfiability entails noncontextuality, hence efficient classical simulation for $M\ge 2$, while with it, satisfiability is essentially automatic.

Let $\rho$ be a state on $L^2(\mathbb R^M)$. Consider the quadruple $(\mathfrak O, \mathcal H,\psi,\pi)$, where $\mathfrak O$ is a von Neumann algebra on the Hilbert space $\mathcal H$, $\psi$ is a unit vector, and $\pi = (\pi_L)_{L\in\operatorname{Lag}(\mathcal X)}$ is as before a family of unital normal $*$-homomorphisms from the context algebras $\mathfrak A_L$ into $\mathfrak O$. We call this a \textit{representation} of the quadrature scenario for $\rho$ if

\begin{enumerate}[\qquad(R1)]
    \item\label{r1} (coherence) $\pi_L(P_{\hat x}(E))=\pi_{L'}(P_{\hat x}(E))$ for Borel $E\subseteq\mathbb{R}$ and $x\in L\cap L'$;
    \item\label{r2} (Born reproduction in $\psi$) $\langle\psi,\pi_L(A)\psi\rangle=\mathrm{Tr}(\rho A)$ for every $L$ and every $A\in\mathfrak{A}_L$.
    \setcounter{continue}{\value{enumi}}
\end{enumerate}

Note that with this definition, quantum mechanics itself provides a representation: for $\psi \in\mathcal H\otimes \mathcal H^\prime$ being the purification of $\rho$, $\mathfrak O=\mathcal L(\mathcal H)\otimes\hat 1$ with $\pi_L(A)=A\otimes \hat 1$ satisfies both (R\ref{r1}) and (R\ref{r2}).

We furthermore call a representation \textit{satisfiable} if
\begin{enumerate}[\qquad(R1)]
    \setcounter{enumi}{\value{continue}}
    \item\label{r3} $\mathcal M^{\mathfrak O}$ is ZFC-satisfiable in $\psi$ in the sense that every $\Delta_0$-formula provable in ZFC holds with probability 1,
    \setcounter{continue}{\value{enumi}}
\end{enumerate}
and \textit{linear} if, as before, 
\begin{enumerate}[\qquad(R1)]
    \setcounter{enumi}{\value{continue}}
    \item\label{r4} $u_xu_y=u_{x+y}$ for all $x,y\in\mathcal{X}$, where $u_x:=\pi_L(e^{i\hat x})$ for any $L\ni x$.
\end{enumerate}

Because of the Weyl relation, the representation furnished by purification within quantum mechanics fails to be linear, and an argument analogous to the one in the proof of Prop.~\ref{prop:nogo} shows it also to be non-satisfiable. 

In the following, we will need the implication from ZFC-satisfiability in $\psi$ to $[X,Y]\psi=0$, (i)$\rightarrow$(iii) in Thm.~\ref{thm:ZFC}. However, in Ref.~\cite{ozawa2023logical}, separability of the Hilbert space is among the assumptions; but this may fail to hold in some cases of the constructions herein, notably $L^\infty(\mathbb R^{\mathcal X},q)$ as used in Prop.~\ref{prop:noncont} (but see the remark on separable alternatives there). While it may be the case that one can always work with separable Hilbert spaces, the following Lemma shows that the implication can be extended to cover the nonseparable case:

\begin{lemma}\label{lem:satcomm}
Let $\mathfrak N$ be a von Neumann algebra on a Hilbert space $\mathcal H$, not necessarily separable, and $\psi\in\mathcal H$ a unit vector. If $\mathcal M^{\mathfrak N}$ is ZFC-satisfiable in $\psi$, then $[X,Y]\psi=0$ for all $X,Y\in\mathfrak N$.
\end{lemma}
\begin{proof}
Let $P,Q$ be projections in $\mathfrak N$ and consider the Boolean-valued sets $u:=\{(\check\emptyset,P)\}$, $v:=\{(\check\emptyset,Q)\}\in \mathcal M^{\mathfrak N}$. Then $\interp{\check\emptyset\in u}=P$, and likewise $\interp{\check\emptyset\in v}=Q$. The quantifier-free formulas
\begin{align}
&\varphi_1(x,y,z):&\ \nonumber\\&x\in y\leftrightarrow\left((x\in y\wedge x\in z)\vee(x\in y\wedge\neg x\in z)\right),\\
&\varphi_2(x,y,z):&\ \nonumber\\&\neg x\in y\leftrightarrow\left((\neg x\in y\wedge x\in z)\vee(\neg x\in y\wedge\neg x\in z)\right)
\end{align}
are tautologies, hence $\Delta_0$ theorems of ZFC. Put $A:=(P\wedge Q)\vee(P\wedge Q^\perp)\le P$ and $B:=(P^\perp\wedge Q)\vee(P^\perp\wedge Q^\perp)\le P^\perp$. Writing $\alpha := (\check\emptyset\in u)$ and $\beta := (\check\emptyset\in v)$, we have $\varphi_1(\check\emptyset, u, v)=\alpha\leftrightarrow((\alpha\wedge\beta)\vee(\alpha\wedge\neg\beta))$, and $\varphi_2$ likewise with $\neg\alpha$ in place of $\alpha$. The truth value of the right hand side of $\varphi_1$ is then exactly $A$, which, with $\interp{\alpha}=P$ entails $\interp{\varphi_1(\check\emptyset, u, v)}=P^\perp\vee A$; similarly, $\interp{\varphi_2(\check\emptyset, u, v)}=P\vee B$. Now, since $A\leq P$, $A$ and $P^\perp$ are orthogonal, so $P^\perp\vee A = P^\perp + A$; likewise, $P\vee B = P + B$. These commute, so their meet is their product:
\begin{align}
    (&P^\perp + A)(P + B)= A + B \nonumber\\ &= (P\wedge Q) + (P\wedge Q^\perp) + (P^\perp\wedge Q) + (P^\perp\wedge Q^\perp)
\end{align}

By satisfiability $\psi$ lies in the range of $A + B$, on which $P$ and $Q$ commute, as $PQ = QP$ on the range of each of the terms of the sum above; thus $[P,Q]\psi=0$. Since $\mathfrak N$ is the norm-closed span of its projections and $(X,Y)\mapsto[X,Y]\psi$ is bilinear and norm-continuous, $[X,Y]\psi=0$ for all $X,Y\in\mathfrak N$.
\end{proof}

The computation of $\interp{\varphi_1(\check\emptyset, u, v)}$ and $\interp{\varphi_2(\check\emptyset, u, v)}$ above needs the evaluation of an implication in an orthomodular lattice, for which multiple options exist. We can follow Ozawa in using the Sasaki hook, $a\rightarrow_S b = a^\perp\vee(a\wedge b)$, but since each biconditional relates commuting projections, any quantum implication that reduces to the Boolean one will yield the same result.

The preceding Lemma now allows us to recast Ozawa's criterion for satisfiability within a state, extended to the possibility of nonseparable Hilbert spaces, as requiring the commutativity of the algebra $\mathfrak O$ restricted to the subspace that can be reached from $\psi$. Take $\mathfrak O$, $\psi$ and $\mathcal H$ as above, and write $\mathfrak O\psi:=\{X\psi:X\in\mathfrak O\}$ and $K:=\overline{\mathfrak O\psi}$ for its norm closure, the \emph{cyclic subspace} generated by $\psi$. Since $\mathfrak O$ is an algebra containing $\hat1$, $K$ is the smallest closed subspace of $\mathcal{H}$ that contains $\psi$ and is invariant under $\mathfrak{O}$.

\begin{lemma}\label{lem:comm}
$\mathcal M^{\mathfrak O}$ is ZFC-satisfiable in $\psi$ if and only if $\mathfrak O_K:=\{X|_K:X\in \mathfrak O\}$ is commutative.
\end{lemma}

\begin{proof}
Since $K$ is invariant under $\mathfrak O$, its projection lies in the commutant $\mathfrak O^\prime$, and the restriction $X\mapsto X|_K$ is a normal unital $*$-homomorphism onto the von Neumann algebra $\mathfrak O_K$ on $K$. In particular, $[X,Y]|_K=[X_K,Y_K]$. For $\mathcal M^{\mathfrak O}$ ZFC-satisfiable in $\psi$, Thm.~\ref{thm:ZFC} resp. its generalization in Lemma~\ref{lem:satcomm} gives $[X,Y]\psi=0$ for all $X,Y\in\mathfrak O$. For three arbitrary operators $X,Y,Z\in\mathfrak O$, applying this to pairs $(X,YZ)$ and $(X,Z)$ yields $XYZ\psi = YZX\psi = YXZ\psi$, so $[X,Y]Z\psi = 0$; thus, $[X,Y]$ vanishes on $\mathfrak O\psi$, and by continuity on $K$, so $\mathfrak O_K$ is commutative. Conversely, if $\mathfrak O_K$ is commutative, $[X,Y]\psi = [X|_K,Y|_K]\psi = 0$ for all $X,Y\in\mathfrak O$, since $\psi\in K$, and Thm.~\ref{thm:ZFC} yields ZFC-satisfiability.
\end{proof}

Note that the converse direction inherits the limitation to separable Hilbert spaces from Thm.~\ref{thm:ZFC}. However, nothing that follows depends on this direction, and it is included here only for completeness.

The following proposition then demonstrates the equivalence between the satisfiability of a representation and noncontextuality.

\begin{proposition}\label{prop:satispsi}
$\rho$ admits a satisfiable representation if and only if it admits a reduced classical system, that is, a triple $(\mathfrak C, p, \pi)$ satisfying (C\ref{c1})--(C\ref{c4}). Thus, by Thm.~\ref{thm:noncont} and Prop.~\ref{prop:noncont}, $\rho$ admits a satisfiable representation if and only if it is noncontextual for quadrature measurements.
\end{proposition}

\begin{proof}
Let $(\mathfrak O, \mathcal H,\psi, \pi)$ be a satisfiable representation and $K:=\overline{\mathfrak O\psi}$ as above. By Lemma~\ref{lem:comm}, $\mathfrak O_K$ is commutative. Since $\psi$ is cyclic for $\mathfrak O_K$, it is separating for $\mathfrak O^\prime_K\supseteq \mathfrak O_K$, so $p:=\left\langle\psi,\cdot\psi \right\rangle$ is a faithful normal state on $\mathfrak O_K$, fulfilling (C\ref{c1}). The maps $\pi_L^K(A):=\pi_L(A)|_K$ are unital normal $*$-homomorphisms $\mathfrak A_L\rightarrow \mathfrak O_K$, satisfying (C\ref{c2}) and (C\ref{c3}) by (R\ref{r1}). Since $\psi\in K$, we have $p(\pi_L^K(A))=\left\langle\psi,\pi_L(A)\psi \right\rangle=\operatorname{Tr}(\rho A)$ by (R\ref{r2}), thus (C\ref{c4}) holds.

Conversely, suppose $(\mathfrak C,p,\pi)$ satisfies (C\ref{c1})--(C\ref{c4}). Let $(\mathcal H_p, \varpi_p,\psi_p)$ be the GNS-representation of $p$. Since $p$ is normal, $\varpi_p(\mathfrak C)$ is a commutative von Neumann algebra on $\mathcal H_p$, so $\mathcal M^{\varpi_p(\mathfrak C)}$ is ZFC-satisfiable in every vector, cf. the transfer principle (Eq.~\eqref{eq:transfer}). The maps $\varpi_p\circ\pi_L$ are unital normal $*$-homomorphisms that satisfy (R\ref{r1}) by (C\ref{c3}), and (R\ref{r2}) holds because $\langle\psi_p,\varpi_p(\pi_L(A))\psi_p\rangle=p(\pi_L(A))=\mathrm{Tr}(\rho A)$.
\end{proof}

Consequently, requiring ZFC-satisfiability in $\psi$ of the representation suffices to pick out the noncontextual scenarios, and via the equivalence to Wigner nonnegativity (in the $M\ge 2$-setting)~\cite{booth2022contextuality,haferkamp2021equivalence}, the ones with an efficient classical simulation. Linearity then corresponds to a stronger condition, ensuring satisfiability in every vector:

\begin{proposition}
\leavevmode
\begin{enumerate}[(i)]
    \item Suppose $(\mathfrak O, \mathcal H,\psi,\pi)$ is a linear, but not necessarily satisfiable, representation for $\rho$. Call $\mathfrak O_0$ the von Neumann algebra generated by $\bigcup_L\pi_L(\mathfrak A_L)$. Then $\mathfrak O_0$ is commutative, and $(\mathfrak O_0, \mathcal H,\psi,\pi)$ with each $\pi_L$ restricted to $\mathfrak O_0$ is a linear satisfiable representation for $\rho$.
    \item The following are equivalent:
    \begin{enumerate}[(a)]
        \item $\rho$ admits a linear representation;
        \item $\rho$ admits a linear satisfiable representation;
        \item $\rho$ admits a classical system fulfilling (C\ref{c1})--(C\ref{c5});
        \item $W_\rho\ge 0$.
    \end{enumerate}
\end{enumerate}
\end{proposition}

\begin{proof}
\begin{enumerate}[(i)]
    \item By the proof of Lemma~\ref{lem:class}(ii), the span of $\{e^{i\hat x}:x\in L\}$ is $\sigma$-weakly dense in $\mathfrak A_L$, so by normality $\pi_L(\mathfrak A_L)$ lies in the von Neumann algebra generated by $\{u_x:x\in L\}$. Hence $\mathfrak O_0$ is generated by $\{u_x:x\in \mathcal X\}$. By linearity (R\ref{r4}), $u_xu_y=u_{x+y}=u_yu_x$, and $u_x^*=u_{-x}$, so this generating set, and with it all of $\mathfrak O_0$, is self-adjoint and commutative. (R\ref{r1}), (R\ref{r2}) and (R\ref{r4}) do not depend on the algebra, and (R\ref{r3}) is fulfilled by commutativity. 
    \item (a)$\Rightarrow$(b) is (i). (b)$\Rightarrow$(c): the construction in the proof of Prop.~\ref{prop:satispsi} yields a triple fulfilling (C\ref{c1})--(C\ref{c4}), and restriction to $K$ preserves (R\ref{r4}) (linearity), which yields (C\ref{c5}). (c)$\Rightarrow$(d) is Thm.~\ref{thm:pos}, and (d)$\Rightarrow$(c) is Prop.~\ref{prop:phase}. (c)$\Rightarrow$(a) then follows from the GNS construction in the proof of Prop.~\ref{prop:satispsi}, turning (C\ref{c5}) into (R\ref{r4}).
\end{enumerate}
\end{proof}

This shows in detail the relation between linearity and ZFC-satisfiability. Imposing linearity makes the algebra generated by the represented contexts commutative, thus ensuring the ZFC-satisfiability of the associated model in every vector. However, perhaps more surprisingly, global commutativity of the algebra is not needed to ensure efficient classical simulation (for $M\ge2$). Imposing Ozawa's notion of ZFC-satisfiability in a state $\psi$ (representing $\rho$ in the sense of reproducing the Born probabilities) yields a necessary and sufficient condition for noncontextuality, giving a precise formulation of the informal discussion in Sec.~\ref{subsec:BVQM}. Both notions differ only for $M=1$: the Fock state $\ket{1}$, for example, admits a satisfiable representation, constructed from the hidden variable model in Sec.~\ref{subsec:phase}, but not a linear one, due to the Wigner function becoming negative near the origin. For $M\ge 2$, they coincide by the results of~\cite{booth2022contextuality, haferkamp2021equivalence}.

The notion of linearity, here, is exactly what was questioned by Hermann~\cite{hermann1935naturphilosophischen} and Bell~\cite{bell1966problem} in von Neumann's claimed proof~\cite{neumann1955mathematical} of the non-existence of hidden variable theories recapitulating the empirical predictions of quantum mechanics. If we thus follow Bell in thinking von Neumann's reasoning `not merely false, but foolish' (as quoted in \cite{mermin1993hidden}), and accordingly are moved to reject linearity, it is exactly ZFC-satisfiability (in $\psi$) that is left as underlying the absence of a quantum advantage. 

\subsection{Relation to Boolean Forcing}

The construction of the classical system has a direct relation to the Boolean forcing constructions described in Sec.~\ref{sec:BVM}. In the continuous variable case, the set of projections $P$ of the algebra $\mathfrak C_\rho = L^\infty(\mathcal X,\mu_\rho)$ are the indicator functions $1_E$ onto the Borel regions $E$ of phase space, identified if they differ only by a $\mu_\rho$-null set. This is a complete Boolean algebra, that is also \textit{atomless}: every region of non-zero probability contains a smaller region of non-zero probability, since $\mu_\rho$ assigns probability zero to individual points. There is thus no smallest non-trivial proposition; no single proposition pins down a phase-space point. Moreover, it is isomorphic to the Lebesgue measure algebra of $[0,1]$, the \textit{random algebra}, by Maharam's theorem~\cite{maharam1942homogeneous} (see also~\cite[§2.1]{boos1996mathematical}). 

Since the Boolean algebra supplying the truth values of the model is effectively the random algebra, the relevant forcing notion is Solovay's \textit{random forcing}~\cite{solovay1970model}, which extends the ground model by a random real number. Starting from a ground model $\mathcal M$, which contains the relevant data such as the measure $\mu_\rho$, we choose a generic ultrafilter $G$. The necessity of genericity has been emphasized by van Wesep~\cite[§3.1]{wesep2006hidden}. This represents a complete and consistent set of answers to all questions of the form `does the phase-space point $\lambda$ lie in $E$?', for all $E$ describable in $\mathcal M$, with the answer being `yes' if and only if the projection onto $E$ is in $G$. From genericity, these answers must be \textit{typical} in the sense that $\lambda$ lies in no region of probability zero that can be described in $\mathcal M$. Thus, $\lambda$ passes every randomness test that can be formulated in $\mathcal M$. However, since it is itself a point, and thus of measure zero, it is not part of the ground model. 

Atomlessness is the condition for forcing to be non-trivial: the resulting forced model is a genuine extension, and non-isomorphic to the ground model. For an atomic Boolean algebra (such as the one arising in the discrete version of the construction, whose atoms are just the phase-space points $v$), any given atom uniquely determines the ultrafilter used in the forcing construction, and that ultrafilter already is a part of the ground model $\mathcal M$. Thus, every consistent set of answers as above just has the form `the point is $v \in S$', so nothing new is added. This setting has a direct connection to the motivating example in Sec.~\ref{sec:BVM}: each atom $v$ defines an individual `world', as the term was used there.

The hidden variable $\lambda$ then specifies all measurement outcomes. Under the Takeuti correspondence, every classical surrogate $x^\flat$ of a quadrature $\hat x$ is a real number of the model. The (projection-valued) truth value of the proposition `$x^\flat\le r$' is the region $\{\eta\in\mathcal X: x\cdot\eta\le r\}$, and its probability equals the Born probability $\mathrm{Tr}\left(\rho\,P_{\hat x}\left((-\infty,r]\right)\right)$. The (classical) truth value of this proposition is then determined by the choice of $G$, i.e. the forcing construction. In $\mathcal M[G]$ the proposition holds if and only if $x\cdot\lambda\le r$, so that the value of $x^\flat$ is $x\cdot\lambda$ for every quadrature label $x$ in $\mathcal M$. Choosing the hidden variable $\lambda$ thus amounts to choosing $G$, which settles all such propositions at once.

Any statement of the form `$x^\flat\leq r$' holds with truth value $1$ (or $0$) iff $W_\rho$ is completely concentrated within that region (or has no probability mass within it at all). Otherwise, the outcome of the associated measurement is logically independent of the information contained within the ground model. This yields a continuous-variable analogue to the observation of Paterek et al.~\cite{paterek2010logical}, already connected to random forcing in~\cite{krol2023random}, that measurements yield definite outcomes for propositions entailed by the axioms encoded in the state, and random outcomes for propositions independent of them.

Each context with its associated algebra $\mathfrak A_L$ can be given the same interpretation. $\mathfrak A_L$ is commutative, and its projections correspond to regions of the outcome space $\mathbb R^M$ of a joint measurement of the quadratures in $L$. The associated model is automatically ZFC-satisfiable (for every state), thus always admits a classical description. The random point added by forcing with this algebra is the joint outcome of the measurement. This kind of reading recapitulates the discussion of Benioff~\cite{benioff1976amodels, benioff1976bmodels}, who considered outcome sequences of infinite strings of repeated measurements. The interpretation of measurement as yielding results via the addition of an infinite random sequence to the ground model was proposed by Król et al.~\cite[Thm. 7]{krol2023random} (cf.~\cite{boos1996mathematical, wesep2006hidden}).

Under this interpretation, Thm.~\ref{thm:pos} can be read as follows. Each context separately admits a beable description, where a joint measurement outcome is determined by a random point on its outcome space. The phase-space system combines all of these into a single beable description: the outcomes of all contexts are the coordinates $x\cdot\lambda$ of one random phase-space point $\lambda$, and the Born probabilities admit an ignorance interpretation as probabilities over the phase-space algebra. This combined description exists precisely when $W_\rho \ge 0$. Consequently, a quantum advantage (for the considered setting) requires that the outcomes of the individual measurement contexts cannot be combined into the coordinates of a single random phase-space point in a beable model. In this sense the quantum advantage requires multiple perspectives.

\section{Conclusion}\label{sec:conc}

We have given a novel sufficient criterion for efficient classical simulation of computations in the continuous-variable and odd-prime dimension cases. This criterion trades on the notion of ZFC-satisfiability: if there exists a beable model in the mathematical-logical sense that carries representations of the quadrature measurements as its internal real numbers, then the computation is efficiently classically simulable (in the sense of~\cite{mari2012positive}) if every ZFC-theorem holds within this model. This gives `ZFC-nonsatisfiability' the character of a \textit{resource} for quantum computing in the continuous variable/qudit setting: a satisfiable model cannot yield a quantum advantage. 

The result was obtained as a bridge between the characterization of contextual hidden variable theories (`beable'-algebras) in terms of the quantum set theory of Ozawa~\cite{ozawa2023logical}, and the equivalence of noncontextuality and Wigner-nonnegativity in the odd-prime dimensional~\cite{delfosse2017equivalence} and continuous cases~\cite{booth2022contextuality, haferkamp2021equivalence}. While the direct approach of finding a beable-subalgebra within the algebra of operators on the Hilbert space of the system (such that the quadratures are included as affiliated operators in the CV case) cannot work, it turns out that an auxiliary construction of mapping the operator algebras for each context into the algebra of (phase-space) functions for the system yields the desired result: a ZFC-satisfiable beable model exists if, and only if, the Wigner function of the system is nonnegative (and the representation is linear, in the case $M=1$). This yields efficient classical simulation by the result of Mari and Eisert~\cite{mari2012positive}. 

Furthermore, we also obtain a logical, state-dependent form of the Kochen-Specker theorem: $\rho$ is contextual if and only if every coherent, Born-reproducing amalgamation of its context models fails to be ZFC-satisfiable in the representing state. There is an interesting connection here to von Neumann's original no-go theorem for hidden variables~\cite{neumann1955mathematical}: his condition is just that of linearity, which picks out exactly the Wigner-nonnegative states. As we have seen, these coincide with the noncontextual states for $M\ge 2$, thus recapturing the Kochen-Specker theorem there; it is only for $M=1$, where the counterexample in Sec.~\ref{subsec:phase} lives, that the two notions diverge. If we thus accept the criticism of Hermann~\cite{hermann1935naturphilosophischen} and Bell~\cite{bell1966problem} of this assumption, it is exactly the notion of ZFC-satisfiability that carries the result.

For the case of $M\ge 2$ qumode systems, efficient classical simulation is secured by the notion of ZFC-satisfiability in a particular state $\psi$, introduced by Ozawa~\cite{ozawa2023logical}. It is there that the logical characterization displays its full consequence: a quantum advantage is possible only if $\rho$ admits no satisfiable representation at all, i.e. if the ZFC-models associated with its individual contexts cannot be merged into a single model that satisfies all ZFC-theorems. 

A necessary condition for the possibility of a quantum advantage in terms of mathematical logic has previously been achieved by van den Nest and Briegel~\cite{van2008measurement}: in the setting of measurement-based quantum computation, they have shown that a quantum advantage is only possible for graph-state families whose underlying graph families have undecidable associated logic theories, making undecidability into a resource in a sense similar to the ZFC-nonsatisfiability considered here. 

Indeed, as already discussed, there is a close connection between ZFC-satisfiability and decidability: Boolean-valued models where sentences assume truth values different from $\{0,1\}$ exist for theories with undecidable statements, which makes possible the existence of inequivalent models of a single theory in the sense that there are sentences $\phi$ that may hold in one model, but fail to do so in another. Thus, in particular, there can be no overarching `view from nowhere' that subsumes all possible local perspectives---the best one can do is Ozawa's orthomodular-valued model $\mathcal M^{\mathbb L}$, which however is no-longer ZFC-satisfiable, such that there exist ZFC-theorems that are assigned a truth value strictly below $1$. In this sense, a quantum advantage needs multiple perspectives, given by local models that cannot be amalgamated into a single `ordinary' global model (where ZFC-theorems have truth value $1$) in order to be possible.

The introduction appealed to undecidability by a heuristic argument, from the randomness of an infinite sequence of measurement outcomes. The empirical relevance of such a notion may be questioned: measurement sequences in the real world are always finite, and any sequence of measurements of finite precision can be mapped to a rational number approximation. However, with the result presented here, the quantum advantage acts as a kind of `herald' of undecidability: if every outcome proposition were decided in $\rho$, the valuations of the individual contexts would agree on overlaps and combine into a single one. Thus, no `second tier' value indefiniteness in the sense of Sec.~\ref{subsec:BVQM} would exist. Indeed, the second tier here is essential: for any noncontextual $\rho$, every undecided proposition is decided within every `world' associated to a single Boolean-valued model, so this first-tier indefiniteness can always be attributed to ignorance about which world obtains. For contextual $\rho$, where ZFC-satisfiability fails in the representing state $\psi$, there is no possible observer, no possible `world' in which every outcome proposition is decided. Thus, a quantum advantage necessitates an \textit{irreducible} undecidability---an epistemic horizon in the sense of \cite{szangolies2018epistemic} that cannot be removed by a simple extension of knowledge. 

One may consider this to vindicate, to some extent, the intuition of Wheeler that the ``point of origin" of the `quantum principle' is ``the `undecidable propositions' of mathematical logic" (see the notes transcribed in the appendix of~\cite{fuchs2016participatory}). If there were no undecidability, then we could always formulate all of the mathematics relevant to the system within a single ZFC-satisfiable model, thus providing us with a notion of classical language. Moreover, it is second-tier value indefiniteness that ensures that quantum probabilities have no ignorance interpretation in general.

While the construction here yields a new \textit{criterion} to judge the presence or absence of a quantum advantage for a given setting, it would be interesting to inquire whether it can perhaps also provide a novel \textit{explanation} of the existence of such an advantage. What is it about the impossibility of representing the mathematics of a given computation within a ZFC-satisfiable model that makes the existence of a quantum advantage possible? Answering this question could yield novel ways of designing quantum algorithms to target exactly this feature in a principled manner, thus allowing a systematic means of producing promising quantum algorithms. 

Bub~\cite{bub2006quantum} (see~\cite{steane2003quantum} for a related perspective) has proposed that the origin of the quantum speedup is to be sought in the non-Boolean event structure of the quantum propositional lattice $\mathbb L(\mathcal H)$. As argued above, this structure becomes relevant exactly once ZFC-satisfiability fails. Essentially, his proposal is that the origin of the speedup lies not in what a quantum computer does, but effectively in what it does \textit{not} have to do: namely, to decide whether a proposition holds, it does not have to assign definite truth-values to every sub-proposition. This has a natural interpretation in terms of Boolean-valued models to the effect that membership within a particular set may have a determinate truth-value without thereby the question of membership within its subsets likewise being decided. A disjunction thus may have value 1 without any of the disjuncts having a determinate value; but a generic extension, determined by a particular ultrafilter, selects definite values also for the disjuncts. However, in the non-ZFC satisfiable case, no such global extension exists, and the disjuncts remain without a definite truth value.

The impossibility of amalgamating different vantage points into a single overarching `view from nowhere' \cite{nagel1986view} has played a prominent role in the resurgence of `perspectivist' approaches to quantum mechanics (see e.g.~\cite{berghofer2026perspectivism} and references therein), spurred in part by renewed interest in extended `Wigner's friend'-type scenarios~\cite{schmid2023review}. The impossibility of finding a universal extension in the non-ZFC satisfiable case, and the necessity to resort to a collection of `local' Boolean-valued models, closely mirrors this kind of perspectivism, and the results presented here tie it to the possibility of a quantum advantage: from within a single perspective, no gain over classical computation is possible. With this in mind, it seems promising to investigate the possibility of a Boolean-valued formulation as providing the foundation of quantum perspectivism writ large, beyond the question of the computational speedup.

Looking in the other direction, the classical systems defined herein might themselves be utilized in efficient simulation of quantum computations, for e.g. benchmarking purposes or in quantum-inspired classical algorithms.

The question remains whether, and how, this result can be extended to the case of quantum computing with qubit systems. Here, the existence of state-independent contextuality, as demonstrated e.g. via the Peres-Mermin argument~\cite{peres1990incompatible, mermin1990quantum}, frustrates a straightforward extension. Consequently, approaches to the necessary resources within this setting have resorted to imposing additional constraints, such as confining the state vector to real values~\cite{delfosse2015wigner} or restricting the allowed set of operations~\cite{raussendorf2017contextuality}. A particularly promising route seems to follow~\cite{thio2026kirkwood} and use the negativity of the Kirkwood-Dirac distribution in the rebit-setting to characterize the possibility of a quantum advantage. If a similar construction to the one presented here can be carried out, this might go some length towards establishing ZFC-satisfiability as a universal sufficient criterion for classical simulability. 

\acknowledgements{
I thank Michael Epping for helpful suggestions and Lawrence B. Crowell for pointing me in the direction of forcing way before I was able to appreciate the hint.

This work received funding from the project ALQU of the DLR Quantum Computing Initiative.

\textbf{Generative AI disclosure statement.} The arguments in Sec.~\ref{sec:qudit}, Sec.~\ref{sec:conti} and Sec.~\ref{sec:logic} were partially developed and/or refined with the assistance of generative AI methods. They were subsequently checked, simplified, and reorganized. The author affirms that he takes full responsibility for the correctness of the results as presented in this paper. What errors remain are fully human.}

\bibliography{BVMSpeedupDOI.bib}

@article{benioff1976amodels,
  title={{Models of Zermelo Frankel Set Theory as Carriers for the Mathematics of Physics. I}},
  author = {Benioff, Paul A.},
  year = {1976},
  month = {May},
  journal = {Journal of Mathematical Physics},
  volume = {17},
  number = {5},
  pages = {618--628},
  issn = {0022-2488, 1089-7658},
  doi = {10.1063/1.522953},
  urldate = {2026-09-04},
  langid = {english}
}

@article{benioff1976bmodels,
  title={{Models of Zermelo Frankel Set Theory as Carriers for the Mathematics of Physics. II}},
  author = {Benioff, Paul A.},
  year = {1976},
  month = {May},
  journal = {Journal of Mathematical Physics},
  volume = {17},
  number = {5},
  pages = {629--640},
  issn = {0022-2488, 1089-7658},
  doi = {10.1063/1.522954},
  urldate = {2026-09-04},
  langid = {english}
}

@inproceedings{takeuti1977boolean,
  title={{Boolean Valued Analysis}},
  author={Takeuti, Gaisi},
  booktitle={{Applications of Sheaves: Proceedings of the Research Symposium on Applications of Sheaf Theory to Logic, Algebra, and Analysis, Durham, July 9--21, 1977}},
  pages={714--731},
  year={1977},
  organization={Springer},
  doi = {10.1007/BFb0061842}
}

@incollection{takeuti1981quantum,
  title={{Quantum Set Theory}},
  author={Takeuti, Gaisi},
  booktitle={{Current Issues in Quantum Logic}},
  pages={303--322},
  year={1981},
  publisher={Springer},
  doi = {10.1007/978-1-4613-3228-2_19}
}

@incollection{landsman2021indeterminism,
  title={{Indeterminism and Undecidability}},
  author={Landsman, Klaas},
  booktitle={{Undecidability, Uncomputability, and Unpredictability}},
  pages={17--45},
  year={2021},
  publisher={Springer},
  doi = {10.1007/978-3-030-70354-7_3}
}

@book{calude2002information,
  title={{Information and Randomness: An Algorithmic Perspective}},
  author={Calude, Cristian S},
  year={2002},
  publisher={Springer Science \& Business Media},
  doi = {10.1007/978-3-662-04978-5}
}

@article{von1963various,
  title={{Various Techniques Used in Connection with Random Digits}},
  author={Von Neumann, John},
  journal={John von Neumann, Collected Works},
  volume={5},
  number={768-770},
  pages={1},
  year={1963},
  publisher={MacMillan New York, USA},
  nolink={}
}

@book{bell2011set,
  title={{Set Theory: Boolean-Valued Models and Independence Proofs}},
  shorttitle = {{Set Theory}},
  author = {Bell, John L.},
  year = 2011,
  month = dec,
  publisher = {Oxford University Press},
  address = {Oxford},
  isbn = {978-0-19-960916-1},
  langid = {Englisch},
  doi={10.1093/acprof:oso/9780198568520.001.0001}
}

@article{cohen1963independence,
  title={{The Independence of the Continuum Hypothesis}},
  author={Cohen, Paul J},
  journal={Proceedings of the National Academy of Sciences},
  volume={50},
  number={6},
  pages={1143--1148},
  year={1963},
  doi = {10.1073/pnas.50.6.1143}
}

@article{szangolies2018epistemic,
  title={{Epistemic Horizons and the Foundations of Quantum Mechanics}},
  author={Szangolies, Jochen},
  journal={Foundations of Physics},
  volume={48},
  number={12},
  pages={1669--1697},
  year={2018},
  publisher={Springer},
  doi = {10.1007/s10701-018-0221-9}
}

@article{kochen1968hidden,
  title={{The Problem of Hidden Variables in Quantum Mechanics}},
  author={Kochen, Simon and Specker, Ernst P},
  journal={Journal of Mathematics and Mechanics},
  number={1},
  volume={17},
  pages={59--87},
  year={1969},
  doi = {10.1512/iumj.1968.17.17004}
}

@article{ozawa2023logical,
  title={{Logical Characterization of Contextual Hidden-Variable Theories Based on Quantum Set Theory}},
  author = {Ozawa, Masanao},
  year = 2023,
  month = nov,
  journal = {Electronic Proceedings in Theoretical Computer Science},
  volume = {394},
  pages = {1--12},
  issn = {2075-2180},
  doi = {10.4204/EPTCS.394.1},
  urldate = {2026-09-04},
  langid = {english}
}

@article{booth2022contextuality,
  title={{Contextuality and Wigner Negativity Are Equivalent for Continuous-Variable Quantum Measurements}},
  author = {Booth, Robert I. and Chabaud, Ulysse and Emeriau, Pierre-Emmanuel},
  year = 2022,
  month = nov,
  journal = {Physical Review Letters},
  volume = {129},
  number = {23},
  pages = {230401},
  issn = {0031-9007, 1079-7114},
  doi = {10.1103/PhysRevLett.129.230401},
  urldate = {2026-09-04},
  langid = {english}
}

@misc{haferkamp2021equivalence,
  title={{Equivalence of Contextuality and Wigner Function Negativity in Continuous-Variable Quantum Optics}},
  author = {Haferkamp, Jonas and {Bermejo-Vega}, Juani},
  year = 2021,
  month = dec,
  number = {arXiv:2112.14788},
  eprint = {2112.14788},
  primaryclass = {quant-ph},
  publisher = {arXiv},
  doi = {10.48550/arXiv.2112.14788},
  urldate = {2026-09-04},
  langid = {english}
}

@article{mari2012positive,
  title={{Positive Wigner Functions Render Classical Simulation of Quantum Computation Efficient}},
  author = {Mari, A. and Eisert, J.},
  year = 2012,
  month = dec,
  journal = {Physical Review Letters},
  volume = {109},
  number = {23},
  publisher = {American Physical Society},
  doi = {10.1103/PhysRevLett.109.230503},
  urldate = {2026-09-07}
}

@misc{bub2006quantum,
  title={{Quantum Computation from a Quantum Logical Perspective}},
  author = {Bub, Jeffrey},
  year = 2006,
  month = oct,
  number = {arXiv:quant-ph/0605243},
  eprint = {quant-ph/0605243},
  publisher = {arXiv},
  doi = {10.48550/arXiv.quant-ph/0605243},
  urldate = {2026-09-04}
}

@article{steane2003quantum,
  title={{A Quantum Computer Only Needs One Universe}},
  author = {Steane, A. M.},
  year = 2003,
  month = sep,
  journal = {Studies in History and Philosophy of Science Part B: Studies in History and Philosophy of Modern Physics},
  volume = {34},
  number = {3},
  eprint = {quant-ph/0003084},
  pages = {469--478},
  issn = {13552198},
  doi = {10.1016/S1355-2198(03)00038-8},
  urldate = {2026-09-04}
}

@book{hoffmann2018grenzen,
  title={{Grenzen der Mathematik: Eine Reise durch die Kerngebiete der mathematischen Logik}},
  shorttitle = {{Grenzen der Mathematik}},
  author = {Hoffmann, Dirk W.},
  year = 2018,
  month = apr,
  publisher = {Springer Spektrum},
  address = {Berlin, Heidelberg},
  isbn = {978-3-662-56617-6},
  langid = {german},
  doi = {10.1007/978-3-662-56617-6}
}

@inproceedings{robinson1950essentially,
  title={{An Essentially Undecidable Axiom System}},
  author={Robinson, Raphael M and others},
  booktitle={Proceedings of the international Congress of Mathematics},
  volume={1},
  pages={729--730},
  year={1950},
  organization={American Mathematical Society 80 Waterman Street, Providence, RI},
  nolink={},
}

@article{godel1931formal,
  title={{{\"U}ber formal unentscheidbare S{\"a}tze der Principia Mathematica und verwandter Systeme I}},
  author={G{\"o}del, Kurt},
  journal={Monatshefte f{\"u}r mathematik und physik},
  volume={38},
  number={1},
  pages={173--198},
  year={1931},
  publisher={Springer},
  doi = {10.1007/BF01700692}
}

@incollection{lawvere2006diagonal,
  title={{Diagonal Arguments and Cartesian Closed Categories}},
  author={Lawvere, F William},
  booktitle={Category Theory, Homology Theory and their Applications II},
  pages={134--145},
  year={1969},
  publisher={Sprin\-ger},
  doi = {10.1007/BFb0080769}
}

@book{takeuti1978two,
  title={{Two Applications of Logic to Mathematics}},
  author = {Takeuti, Gaisi},
  year = 1978,
  publisher = {Princeton University Press},
  doi = {10.1515/9781400871346}
}

@article{clifton1996properties,
  title={{The Properties of Modal Interpretations of Quantum Mechanics}},
  author={Clifton, Rob},
  journal={The British journal for the philosophy of science},
  volume={47},
  number={3},
  pages={371--398},
  year={1996},
  publisher={Oxford University Press},
  doi = {10.1093/bjps/47.3.371}
}

@article{boos1996mathematical,
  title={{Mathematical Quantum Theory I: Random Ultrafilters as Hidden Variables}},
  author = {Boos, William},
  year = 1996,
  journal = {Synthese},
  volume = {107},
  number = {1},
  pages = {83--143},
  issn = {0039-7857, 1573-0964},
  doi = {10.1007/BF00413903}
}

@article{wesep2006hidden,
  title={{Hidden Variables in Quantum Mechanics: Generic Models, Set-Theoretic Forcing, and the Emergence of Probability}},
  shorttitle = {Hidden Variables in Quantum Mechanics},
  author = {Wesep, Robert A. Van},
  year = 2006,
  journal = {Annals of Physics},
  volume = {321},
  number = {10},
  eprint = {quant-ph/0506040},
  pages = {2453--2475},
  issn = {00034916},
  doi = {10.1016/j.aop.2006.02.003}
}

@article{birkhoff1936logic,
 author = {Garrett Birkhoff and John Von Neumann},
 journal = {Annals of Mathematics},
 number = {4},
 pages = {823--843},
 publisher = {[Annals of Mathematics, Trustees of Princeton University on Behalf of the Annals of Mathematics, Mathematics Department, Princeton University]},
 title={{The Logic of Quantum Mechanics}},
 urldate = {2026-09-09},
 volume = {37},
 year = {1936},
 doi = {10.2307/1968621}
}

@article{davis1977relativity,
  title={{A Relativity Principle in Quantum Mechanics}},
  author = {Davis, Martin},
  year = 1977,
  month = nov,
  journal = {International Journal of Theoretical Physics},
  volume = {16},
  number = {11},
  pages = {867--874},
  issn = {0020-7748, 1572-9575},
  doi = {10.1007/BF01807619},
  urldate = {2026-09-04},
  copyright = {http://www.springer.com/tdm},
  langid = {english}
}

@article{ozawa2017orthomodular,
  title={{Orthomodular-Valued Models for Quantum Set Theory}},
  author = {Ozawa, Masanao},
  year = 2017,
  month = dec,
  journal = {The Review of Symbolic Logic},
  volume = {10},
  number = {4},
  pages = {782--807},
  issn = {1755-0203, 1755-0211},
  doi = {10.1017/S1755020317000120}
}

@article{ozawa2021quantum,
  title={{Quantum Set Theory: Transfer Principle and De Morgan's Laws}},
  shorttitle = {Quantum {{Set Theory}}},
  author = {Ozawa, Masanao},
  year = 2021,
  month = apr,
  journal = {Annals of Pure and Applied Logic},
  volume = {172},
  number = {4},
  pages = {102938},
  issn = {01680072},
  doi = {10.1016/j.apal.2020.102938}
}

@article{ozawa2007transfer,
  title={{Transfer Principle in Quantum Set Theory}},
  author={Ozawa, Masanao},
  journal={The Journal of Symbolic Logic},
  volume={72},
  number={2},
  pages={625--648},
  year={2007},
  publisher={Cambridge University Press},
  doi = {10.2178/jsl/1185803627}
}

@article{krol2025negation,
  title={{Negation of the Smooth Poincare Conjecture in Dimension 4 and Negation of the Tsirelson's Conjecture Shed Light on Quantum Gravity}},
  author = {Kr{\'o}l, Jerzy and {Asselmeyer-Maluga}, Torsten},
  year = 2025,
  month = apr,
  journal = {Universe},
  volume = {11},
  number = {4},
  pages = {126},
  issn = {2218-1997},
  doi = {10.3390/universe11040126}
}

@article{halvorson1999maximal,
  title={{Maximal Beable Subalgebras of Quantum Mechanical Observables}},
  author = {Halvorson, Hans and Clifton, Rob},
  year = 1999,
  month = oct,
  journal = {International Journal of Theoretical Physics},
  volume = {38},
  number = {10},
  pages = {2441--2484},
  issn = {1572-9575},
  doi = {10.1023/A:1026628407645}
}

@book{bell2004speakable,
  title={{Speakable and Unspeakable in Quantum Mechanics: Collected Papers on Quantum Philosophy}},
  author={Bell, John Stewart},
  year={2004},
  publisher={Cambridge university press},
  doi = {10.1017/CBO9780511815676}
}

@article{krol2023random,
  title={{Random World and Quantum Mechanics}},
  author = {Kr{\'o}l, Jerzy and Bielas, Krzysztof and {Asselmeyer-Maluga}, Torsten},
  year = 2023,
  month = jun,
  journal = {Foundations of Science},
  volume = {28},
  number = {2},
  pages = {575--625},
  issn = {1233-1821, 1572-8471},
  doi = {10.1007/s10699-022-09852-2}
}

@article{paterek2010logical,
  title={{Logical Independence and Quantum Randomness}},
  author={Paterek, Tomasz and Kofler, Johannes and Prevedel, Robert and Klimek, Peter and Aspelmeyer, Markus and Zeilinger, Anton and Brukner, {\v{C}}aslav},
  journal={New Journal of Physics},
  volume={12},
  number={1},
  pages={013019},
  year={2010},
  doi = {10.1088/1367-2630/12/1/013019}
}

@article{einstein1935can,
  title={{Can Quantum-Mechanical Description of Physical Reality Be Considered Complete?}},
  author={Einstein, Albert and Podolsky, Boris and Rosen, Nathan},
  journal={Physical review},
  volume={47},
  number={10},
  pages={777},
  year={1935},
  publisher={APS},
  doi = {10.1103/PhysRev.47.777}
}

@incollection{halvorson2002reconsidering,
  title={{Reconsidering Bohr’s Reply to EPR}},
  author={Halvorson, Hans and Clifton, Rob},
  booktitle={Non-locality and Modality},
  pages={3--18},
  year={2002},
  publisher={Springer},
  doi = {10.1007/978-94-010-0385-8_1}
}

@phdthesis{howard1979complementarity,
  title={{Complementarity and Ontology: Niels Bohr and the Problem of Scientific Realism in Quantum Physics}},
  author={Howard, Don Armin},
  type={PhD thesis},
  year={1979},
  school={Boston University Graduate School}
}

@article{barbosa2022continuous,
  title={{Continuous-Variable Nonlocality and Contextuality}},
  author={Barbosa, Rui Soares and Douce, Tom and Emeriau, Pierre-Emmanuel and Kashefi, Elham and Mansfield, Shane},
  journal={Communications in Mathematical Physics},
  volume={391},
  number={3},
  pages={1047--1089},
  year={2022},
  publisher={Springer},
  doi = {10.1007/s00220-021-04285-7}
}

@book{schmudgen2012unbounded,
  title={{Unbounded Self-Adjoint Operators on Hilbert Space}},
  author={Schm{\"u}dgen, Konrad},
  year={2012},
  publisher={Springer Science \& Business Media},
  doi = {10.1007/978-94-007-4753-1}
}

@book{kadison1986fundamentals,
  title={{Fundamentals of the Theory of Operator Algebras. Volume II: Advanced Theory}},
  author={Kadison, Richard V and Ringrose, John R},
  volume={25},
  year={1986},
  publisher={Academic press New York},
  doi={10.1090/gsm/016}
}

@book{cohn2013measure,
  title={{Measure Theory}},
  author={Cohn, Donald L},
  volume={2},
  year={2013},
  publisher={Springer},
  doi = {10.1007/978-1-4614-6956-8}
}

@article{van2008measurement,
  title={{Measurement-Based Quantum Computation and Undecidable Logic}},
  author={Van den Nest, Maarten and Briegel, Hans J},
  journal={Foundations of Physics},
  volume={38},
  number={5},
  pages={448--457},
  year={2008},
  publisher={Springer},
  doi = {10.1007/s10701-008-9212-6}
}

@incollection{fuchs2016participatory,
  title={{On Participatory Realism}},
  author={Fuchs, Christopher A},
  booktitle={Information and Interaction: Eddington, Wheeler, and the Limits of Knowledge},
  pages={113--134},
  year={2016},
  publisher={Springer},
  doi = {10.1007/978-3-319-43760-6_7}
}

@article{raussendorf2017contextuality,
  title={{Contextuality and Wigner-Function Negativity in Qubit Quantum Computation}},
  author={Raussendorf, Robert and Browne, Dan E and Delfosse, Nicolas and Okay, Cihan and Bermejo-Vega, Juan},
  journal={Physical Review A},
  volume={95},
  number={5},
  pages={052334},
  year={2017},
  publisher={APS},
  doi = {10.1103/PhysRevA.95.052334}
}

@article{howard2014contextuality,
  title={{Contextuality Supplies the `Magic' for Quantum Computation}},
  author = {Howard, Mark and Wallman, Joel and Veitch, Victor and Emerson, Joseph},
  year = 2014,
  month = jun,
  journal = {Nature},
  volume = {510},
  number = {7505},
  pages = {351--355},
  publisher = {Nature Publishing Group},
  issn = {1476-4687},
  doi = {10.1038/nature13460}
}

@article{delfosse2017equivalence,
  title={{Equivalence between Contextuality and Negativity of the Wigner Function for Qudits}},
  author = {Delfosse, Nicolas and Okay, Cihan and {Bermejo-Vega}, Juan and Browne, Dan E and Raussendorf, Robert},
  year = 2017,
  month = dec,
  journal = {New Journal of Physics},
  volume = {19},
  number = {12},
  pages = {123024},
  publisher = {IOP Publishing},
  issn = {1367-2630},
  doi = {10.1088/1367-2630/aa8fe3}
}

@article{gross2006hudson,
  title={{Hudson’s Theorem for Finite-Dimensional Quantum Systems}},
  author={Gross, David},
  journal={Journal of mathematical physics},
  volume={47},
  number={12},
  year={2006},
  publisher={AIP Publishing},
  doi = {10.1063/1.2393152}
}

@article{delfosse2015wigner,
  title={{Wigner Function Negativity and Contextuality in Quantum Computation on Rebits}},
  author={Delfosse, Nicolas and Allard Guerin, Philippe and Bian, Jacob and Raussendorf, Robert},
  journal={Physical Review X},
  volume={5},
  number={2},
  pages={021003},
  year={2015},
  publisher={APS},
  doi = {10.1103/PhysRevX.5.021003}
}

@article{peres1990incompatible,
  title={{Incompatible Results of Quantum Measurements}},
  author={Peres, Asher},
  journal={Physics Letters A},
  volume={151},
  number={3-4},
  pages={107--108},
  year={1990},
  publisher={Elsevier},
  doi = {10.1016/0375-9601(90)90172-K}
}

@article{mermin1990quantum,
  title={{Simple Unified Form for the Major No-Hidden-Variables Theorems}},
  author={Mermin, N David},
  journal={Physical Review Letters},
  volume={65},
  number={27},
  pages={3373},
  year={1990},
  publisher={APS},
  doi = {10.1103/PhysRevLett.65.3373}
}

@article{thio2026kirkwood,
  title={{Kirkwood-Dirac Nonpositivity Is a Necessary Resource for Quantum Computing}},
  author={Thio, Jonathan J and Yang, Songqinghao and Yunger Halpern, Nicole and De Bievre, Stephan and Barnes, Crispin HW and Arvidsson-Shukur, David RM},
  journal={Physical Review Letters},
  volume={137},
  number={8},
  pages={080601},
  year={2026},
  publisher={APS},
  doi = {10.1103/x819-898d}
}

@book{neumann1955mathematical,
    title={{Mathematical Foundations of Quantum Mechanics}},
    author={von Neumann, John},
    year={1955},
    publisher={Princeton University Press},
    address={Princeton},
    doi={10.1515/9781400889921}
}

@article{hermann1935naturphilosophischen,
  title={{Die Naturphilosophischen Grundlagen der Quantenmechanik}},
  author={Hermann, Grete},
  journal={Naturwissenschaften},
  volume={23},
  number={42},
  pages={718--721},
  year={1935},
  publisher={Springer-Verlag, Berlin/Heidelberg},
  doi = {10.1007/BF01491142}
}

@book{crull2016grete,
  title={{Grete Hermann -- Between Physics and Philosophy}},
  author={Crull, Elise and Bacciagaluppi, Guido},
  year={2016},
  publisher={Springer},
  doi = {10.1007/978-94-024-0970-3}
}

@article{bell1966problem,
  title={{On the Problem of Hidden Variables in Quantum Mechanics}},
  author={Bell, John S},
  journal={Reviews of Modern physics},
  volume={38},
  number={3},
  pages={447},
  year={1966},
  publisher={APS},
  doi = {10.1103/RevModPhys.38.447}
}

@article{bub2010neumann,
  title={{Von Neumann’s ‘No Hidden Variables’ Proof: A Re-appraisal}},
  author={Bub, Jeffrey},
  journal={Foundations of Physics},
  volume={40},
  number={9},
  pages={1333--1340},
  year={2010},
  publisher={Springer},
  doi = {10.1007/s10701-010-9480-9}
}

@article{dieks2017neumann,
  title={{Von Neumann’s Impossibility Proof: Mathematics in the Service of Rhetorics}},
  author={Dieks, Dennis},
  journal={Studies in History and Philosophy of Science Part B: Studies in History and Philosophy of Modern Physics},
  volume={60},
  pages={136--148},
  year={2017},
  publisher={Elsevier},
  doi = {10.1016/j.shpsb.2017.01.008}
}

@article{abramsky2011sheaf,
  title={{The Sheaf-Theoretic Structure of Non-locality and Contextuality}},
  author={Abramsky, Samson and Brandenburger, Adam},
  journal={New Journal of Physics},
  volume={13},
  number={11},
  pages={113036},
  year={2011},
  publisher={IOP Publishing},
  doi = {10.1088/1367-2630/13/11/113036}
}

@article{isham1998topos,
  title={{Topos Perspective on the Kochen-Specker Theorem: I. Quantum States as Generalized Valuations}},
  author={Isham, Chris J and Butterfield, Jeremy},
  journal={International journal of theoretical physics},
  volume={37},
  number={11},
  pages={2669--2733},
  year={1998},
  publisher={Springer},
  doi = {10.1023/A:1026680806775}
}

@article{doring2021bridge,
  title={{A Bridge between Q-Worlds}},
  author={D{\"o}ring, Andreas and Eva, Benjamin and Ozawa, Masanao},
  journal={The Review of Symbolic Logic},
  volume={14},
  number={2},
  pages={447--486},
  year={2021},
  publisher={Cambridge University Press},
  doi = {10.1017/S1755020319000492}
}

@article{maharam1942homogeneous,
  title={{On Homogeneous Measure Algebras}},
  author={Maharam, Dorothy},
  journal={Proceedings of the National Academy of Sciences},
  volume={28},
  number={3},
  pages={108--111},
  year={1942},
  doi = {10.1073/pnas.28.3.108}
}

@article{solovay1970model,
  title={{A Model of Set-Theory in Which Every Set of Reals Is Lebesgue Measurable}},
  author={Solovay, Robert M},
  journal={Annals of Mathematics},
  volume={92},
  number={1},
  pages={1--56},
  year={1970},
  publisher={JSTOR},
  doi = {10.2307/1970696}
}

@article{veitch2012negative,
  title={{Negative Quasi-probability as a Resource for Quantum Computation}},
  author={Veitch, Victor and Ferrie, Christopher and Gross, David and Emerson, Joseph},
  journal={New Journal of Physics},
  volume={14},
  number={11},
  pages={113011},
  year={2012},
  publisher={IOP Publishing},
  doi = {10.1088/1367-2630/14/11/113011}
}

@article{bravyi2005universal,
  title={{Universal Quantum Computation with Ideal Clifford Gates and Noisy Ancillas}},
  author={Bravyi, Sergey and Kitaev, Alexei},
  journal={Physical Review A—Atomic, Molecular, and Optical Physics},
  volume={71},
  number={2},
  pages={022316},
  year={2005},
  publisher={APS},
  doi = {10.1103/PhysRevA.71.022316}
}

@article{mermin1993hidden,
  title={{Hidden Variables and the Two Theorems of John Bell}},
  author={Mermin, N David},
  journal={Reviews of Modern Physics},
  volume={65},
  number={3},
  pages={803},
  year={1993},
  publisher={APS},
  doi = {10.1103/RevModPhys.65.803}
}

@article{hudson1974wigner,
  title={{When is the Wigner Quasi-Probability Density Non-Negative?}},
  author={Hudson, Robin L},
  journal={Reports on mathematical physics},
  volume={6},
  number={2},
  pages={249--252},
  year={1974},
  publisher={Elsevier},
  doi = {10.1016/0034-4877(74)90007-X}
}

@article{soto1983wigner,
  title={When is the Wigner Function of Multidimensional Systems Nonnegative?},
  author={Soto, Francisco and Claverie, Pierre},
  journal={Journal of Mathematical Physics},
  volume={24},
  number={1},
  pages={97--100},
  year={1983},
  publisher={American Institute of Physics},
  doi = {10.1063/1.525607}
}

@article{bu2025extremality,
  title={Extremality of stabilizer states},
  author={Bu, Kaifeng},
  journal={Physical Review A},
  volume={112},
  number={3},
  pages={032420},
  year={2025},
  publisher={APS},
  doi = {10.1103/ln8j-ytcj}
}

@book{nagel1986view,
  title={{The View from Nowhere}},
  author={Nagel, Thomas},
  year={1986},
  publisher={Oxford University Press},
  address={Oxford},
  doi={10.2307/2220404}
}

@article{berghofer2026perspectivism,
  title={{Perspectivism and Quantum Mechanics}},
  author={Berghofer, Philipp and Khalili, Mahdi},
  journal={Foundations of Physics},
  volume={56},
  number={4},
  pages={36},
  year={2026},
  publisher={Springer},
  doi={10.1007/s10701-026-00937-7}
}

@misc{schmid2023review,
  title={{A Review and Analysis of Six Extended Wigner's Friend Arguments}}, 
  author={David Schmid and Y\`il\`e Y\=ing and Matthew Leifer},
  year={2024},
  eprint={2308.16220},
  archivePrefix={arXiv},
  primaryClass={quant-ph},
  url={https://arxiv.org/abs/2308.16220}, 
  doi={10.48550/arXiv.2308.16220}
}

\end{document}